# Spin Gauge Fields: from Berry Phase to Topological Spin Transport and Hall Effects

K.Yu. Bliokh[1,2*], Yu.P. Bliokh[3]

[1]*Institute of Radio Astronomy, 4 Krasnoznamyonnaya st., Kharkov, 61002, Ukraine*
[2]*Department of Physics, Bar-Ilan University, Ramat-Gan, 52900, Israel*
[3]*Department of Physics, Technion, Haifa, 32000, Israel*

The paper examines the emergence of gauge fields during the evolution of a particle with a spin that is described by a matrix Hamiltonian with $n$ different eigenvalues. It is shown that by introducing a spin gauge field a particle with a spin can be described as a spin multiplet of scalar particles situated in a non-Abelian pure gauge (forceless) field $\mathbf{U}(n)$. As the result, one can create a theory of particle evolution that is gauge invariant with regards to the group $\mathbf{U}^n(1)$. Due to this, in the adiabatic (Abelian) approximation the spin gauge field is an analogue of $n$ electromagnetic fields $\mathbf{U}(1)$ on the extended phase space of the particle. These fields are force ones, and the forces of their action enter the particle motion equations that are derived in the paper in the general form. The motion equations describe the topological spin transport, pumping and splitting. The Berry phase is represented in this theory analogously to the Dirac phase of a particle in an electromagnetic field. Due to the analogy with the electromagnetic field, the theory becomes natural in the four-dimensional form. Besides the general theory the article considers a number of important particular examples, both known and new.

## 1. INTRODUCTION AND BASIC IDEAS

**1.1. A charge in an external electromagnetic field and the Dirac phase.** Let's recall the basic facts about external electromagnetic field influence upon a charged particle. These properties will be of use when constructing the analogous theory for other fields. The classical electromagnetic field is specified by the 4-potential $\mathscr{A}^\alpha = \mathscr{A}^\alpha(r^\alpha)$, which corresponds to the field strength tensor

$$\mathscr{F}^{\alpha\beta} = \frac{\partial \mathscr{A}^\beta}{\partial r_\alpha} - \frac{\partial \mathscr{A}^\alpha}{\partial r_\beta} \ . \tag{1.1}$$

(Throughout the paper we denote the electromagnetic field components and the potential components by script letters; the values equipped with Greek indices are 4-vectors in the Minkowski space with the signature $(-+++)$, a summation over repeated indices is understood.) The electromagnetic field tensor is an invariant with respect to the gauge transformation of the potential:

$$\mathscr{A}^\alpha \to \mathscr{A}^\alpha - \frac{\hbar c}{e}\frac{\partial \phi}{\partial r_\alpha} \ . \tag{1.2}$$

Here, $\phi = \phi(r^\alpha)$ is an arbitrary scalar field on $r^\alpha$, while the coefficient $\hbar c/e$ has been introduced for convenience ($e$ stands for the electric charge of the particle).

---

[*] E-mail: k_bliokh@mail.ru

The force action of the electromagnetic field on a classical charged particle is reduced to the occurrence of the force $\mathscr{F}^{\alpha\beta}\dot{r}_\beta$ in the equations of motion (see Subsection 3.3 below). At the same time, the proper potential $\mathscr{A}^\alpha$ in no way affects the evolution of a classical particle. The situation is different for a quantum particle in an external electromagnetic field.

In 1931, P.A.M. Dirac [1] showed that when the particle is being transported in an external electromagnetic field, the phase $\varphi$ of its wave function $\psi$ contains the term

$$\varphi_D = \frac{e}{\hbar c}\int \mathscr{A}^\alpha dr_\alpha \ . \qquad (1.3)$$

The phase $\varphi_D$ is said to be the *Dirac phase*. (It is evident that with the particle evolution, in addition to phase (1.3), the usual, dynamic phase accumulates.) To derive formula (1.3), Dirac invoked the *gauge invariance* property of the quantum-mechanical equations. This property implies that the equations are invariant with respect to the local gauge transformation of the potential (1.2) and simultaneous multiplication of the particle wave function by the corresponding phase factor

$$\mathscr{A}^\alpha \to \mathscr{A}^\alpha - \frac{\hbar c}{e}\frac{\partial \phi}{\partial r_\alpha} \ , \ \psi \to \psi \exp(i\phi) \ . \qquad (1.2a)$$

The Dirac phase (1.3) is *nonintegrable*, i.e. is dependent on the transport path. Under the transfer along a closed contour, the change of the phase is determined by the field flux through this contour. Indeed, with the charge transfer along a loop that bounds an element of the surface $s^{\alpha\beta}$, the equation (1.3), according to the Stokes theorem, takes the form

$$\varphi_D = \frac{e}{\hbar c}\oint \mathscr{A}^\alpha dr_\alpha = \frac{e}{\hbar c}\int \mathscr{F}^{\alpha\beta} ds_{\alpha\beta} \ . \qquad (1.4)$$

In the magnetic field $\mathscr{B}$ (with no electric field), in the case of a loop in three-dimensional $\mathbf{r}$-space, equation (1.4) is considered without the time component. Then, the vector-potential $\mathscr{A}$ plays the role of $\mathscr{A}^\alpha$, whereas the magnetic field flux $\Phi$ through the loop acts as the flux:

$$\varphi_D = \frac{e}{\hbar c}\oint \mathscr{A} d\mathbf{r} = \frac{e}{\hbar c}\int \mathscr{B} d\mathbf{s} = \frac{e}{\hbar c}\Phi \ . \qquad (1.4a)$$

Phase (1.4), (1.4a) is gauge invariant with respect to transformations (1.2), and hence, may give rise to observable physical effects.

In 1959, the Aharonov-Bohm effect [2] (see also [11]), closely related to the Dirac phase (1.3), (1.4) was revealed. This effect is purely quantum in nature; it consists in the following: a particle can experience an electromagnetic field influence in the region where the field strength $\mathscr{F}^{\alpha\beta}$ is identically zero, while the potential $\mathscr{A}^\alpha$ is nonvanishing. The simplest model demonstrating the Aharonov-Bohm effect is the electron scattering by an infinite thin solenoid containing a flux of a magnetic field. Indeed, outside of the solenoid, the electromagnetic field equals zero, whereas the potential is nonzero, because the contour integral $\oint \mathscr{A} d\mathbf{r}$ of the potential over the loop enclosing the solenoid should be equal to the nonvanishing magnetic flux in the solenoid. When considering the electron scattering by a solenoid, account must be taken of the quantum interference effects of all possible electron trajectories. As a result, the trajectories bypassing the solenoid from opposite sides will acquire the Dirac phase difference (1.3), which can be given in the form of Eq. (1.4a) in terms of the magnetic flux in the solenoid. It is this interference effect that is observed experimentally. Besides, the same effect causes the *deviation* of the narrow electron beam during its scattering on an infinite thin solenoid [3].

In the contemporary geometric language various fields are described as *gauge fields* (see, for example, [4]). At that, the gauge potential is the *connection* in the principal bundle over a space $r^\alpha$ with some structural group $\mathbf{G}$, while a tensor of the gauge field is the tensor of the *curvature* associated with this connection. In this way, the electromagnetic field is the gauge field with the structural group $\mathbf{U}(1)$. The structural gauge group describes the symmetry of the equations with respect to corresponding local gauge transformations of the wave function $\psi \to g(r^\alpha)\psi$ ($g \in \mathbf{G}$) (the second equation (1.2a) in the case of an electromagnetic field). The gauge field is introduced exactly to make the particle motion equation invariant under these transformations. This gauge field in its turn is also transformed in a certain way (the first equation (1.2a) for electromagnetic field). Thus, if the potential $\mathscr{A}^\alpha$ determines the connection in the principal bundle, expressions (1.3) and (1.4) for the Dirac phase define nothing but the law of the *parallel transport* of the vector. Obviously, the phase $\varphi$ of the wave function represents a one-dimensional vector in the bundle layer where the group $\mathbf{U}(1)$ acts.

T.T. Wu and Ch.N. Yang have shown in [5] that one can use the corresponding phase factor of the wave function to completely describe the gauge field. Thus, all electromagnetic phenomena can be derived from the Dirac phase (1.3), (1.4). In other words, quoting [5], *electromagnetism is the gauge-invariant manifestation of the nonintegrable Dirac phase factor*. This kind of the 'phase' approach to the gauge field theory will be of use below.

**1.2. Hermitian Hamiltonian on the parameter space and Berry phase.** In 1984 M.V. Berry made an interesting discovery [6]. He studied the transport of the eigenstate vector of a particle described by the Hermitian matrix Hamiltonian $\hat{H}$. (Here and further we equip all the values associated with the matrices with caps.) It was assumed that the Hamiltonian depended on several independent parameters $\boldsymbol{\mu}$, $\hat{H} = \hat{H}(\boldsymbol{\mu})$, which in their turn depended somehow on independent variables (time, for example). Thus, the problem of the Hamiltonian eigenvector transport was analyzed in the $\boldsymbol{\mu}$-space of the parameters. (Running ahead, it should be noted that ordinary coordinates, the momentum of the particle, as well as some abstract parameters might act in various problems as the $\boldsymbol{\mu}$ parameters.) The essence of Berry's discovery is that the space of parameters considered was found to possess nontrivial connection $\hat{\mathbf{A}} = \hat{\mathbf{A}}(\boldsymbol{\mu})$ under the parallel transport of the Hamiltonian eigenvectors [7,8]. This fact has resulted in the nonintegrable term in the particle's phase, which is similar to the Dirac phase (1.3), (1.4), and (1.4a) with the connection $\hat{\mathbf{A}}$:

$$\varphi_B = \int \hat{\mathbf{A}} \, d\boldsymbol{\mu} . \qquad (1.5)$$

The analogy with the Dirac phase is quite natural since the normalized eigenvector of the Hermitian matrix is determined up to an arbitrary phase factor $\exp(i\phi)$, that is, it possesses the local gauge symmetry $\mathbf{U}(1)$ (see Remark 1 below), like the wave function of a particle in an electromagnetic field does. The nonintegrable phase that the eigenvectors of the Hermitian Hamiltonian possess is called *Berry phase* or *the geometric (topological) phase*. Today this phenomenon is well studied theoretically and repeatedly observed experimentally (see, for example, [9–15] and References there).

It turns out that the nontrivial connection $\hat{\mathbf{A}}$ occurs in the parameter space due to the presence of the *points of degeneracy* (term intersections) of the eigenvalues of the Hamiltonian in question. These points of degeneracy act as the analogs of *magnetic monopoles* (or *topological charges* with the Chern numbers) during independent (adiabatic) evolution of the particle eigenstates [6–9], see also [62]. Hence, if the degeneracy and the contour, along which the particle moves in the $\boldsymbol{\mu}$-space, do not lie in the same plane (which is a generic case), then the non-zero analog of the magnetic flux through this contour exists, and the eigenstate wave function acquires the analogue to the Dirac phase (1.4a). It is significant that in the generic case of the Hermitian Hamiltonian, the $\boldsymbol{\mu}$-space, in which the degeneracies appear as points, is 3-dimensional [6–8]. This result goes back to Von Neumann and Wigner's paper [16]. The curvature $\hat{\mathbf{F}} = \hat{\mathbf{F}}(\boldsymbol{\mu})$ associated with Berry's connection $\hat{\mathbf{A}}$ turns out to be nonzero in the adiabatic approximation and occurs in the expression for the Berry phase for loops in the $\boldsymbol{\mu}$-space in the same manner as the magnetic field $\mathscr{B}$ does in Eq. (1.4a):

$$\varphi_B^{(ad)} = \oint \hat{\mathbf{A}}^{(ad)} d\boldsymbol{\mu} = \int \hat{\mathbf{F}}^{(ad)} d\boldsymbol{\sigma} . \qquad (1.6)$$

Here $d\boldsymbol{\sigma}$ is the oriented element of area in the $\boldsymbol{\mu}$-space and the superscripts (*ad*) stress that this formula is true in adiabatic approximation only (see below). (It will be shown below that in adiabatic approximation for Hamiltonian without global degenerations all matrix quantities in Eq. (1.6) are diagonal. Theirs different eigen values correspond to different eigen states of the particle.) As mentioned above, in this situation the configuration of $\hat{\mathbf{F}}^{(ad)}$ corresponds to the field of the magnetic monopole; in other words, it has the Coulomb-type singularities at the degeneracies in the generic case.

**1.3. The spin gauge field.** The analogy between the notions of the connection and the gauge field leads immediately to the idea of treating the above-considered nontrivial Berry's connection $\hat{\mathbf{A}}$ as some gauge potential. At the same time, the presence of some nonzero field tensor (curvature) $\hat{\mathbf{F}}^{(ad)}$ must not only affect the particle phase but also have a direct *force* action or have an influence on the particle *transport*.

Indeed, just after the discovery of the Berry phase the manifestations of the so-called *geometric force* were revealed [17–26]. And almost simultaneously with the discovering of the Berry phase D.J. Thouless proposed a mechanism of adiabatic transport of particles [26]. Recently this lead to discovery and intensive research of *charge adiabatic pumping* [27–40], and later *spin adiabatic pumping* [41–53]. At the same time the authors of the works [54–84] described the emergence of the charge and spin *dissipationless currents* of particles in various solids. Now it is clear that all the mentioned phenomena share the same nature (that is connected to the Berry phase and corresponding gauge fields) and can be ascribed to the *topological spin transport* of particles (the charge transport is simply the sum of the spin transports over all spin states). The difference between manifestations of the geometric forces and adiabatic pumping [17–53] on one hand and dissipationless currents, which are considered in [54–84], on the other amounts just to that the gauge field acts in the former case on the real space and time, while in the latter case it acts on the momentum space. The commonness of these cases is shown in papers [47,55,76]. Topological spin transport is investigated now also for relativistic particles in the Dirac equation framework [86] and for photons in geometrical optics approximation [87–91].

The topic of topological spin transport attracts now an intense interest. It is supposed that the mentioned effects will assist in controlling and directing the evolution of various spin states of particles and play a great role in spintronics. The developed fragments of the theory of the particle motion in the gauge fields $\hat{\mathbf{A}}^{(ad)}$ and $\hat{\mathbf{F}}^{(ad)}$ have immediately explained various experimentally observed physical phenomena such as the anomalous Hall effect [56–58,62,63,69,71,72,81], electric polarization of solids [39], and the optical Magnus effect [86–91,86], as well as predict novel phenomena like spin Hall effect [58,59,64–68,70,73–75,77–80,82] and topological spin splitting [47,84,87–90]. All of these phenomena are nothing but manifestations of Berry's gauge fields $\hat{\mathbf{A}}$ and $\hat{\mathbf{F}}$ in the equations of motion and evolution of particles.

The Berry phase and the fields $\hat{\mathbf{A}}$ and $\hat{\mathbf{F}}$ are frequently associated with the particle *spin*. The point is that these phenomena arise only if the particle is described by a *vector* (*multicomponent*) wave function, while the Hamiltonian is accordingly given by the Hermitian matrix. The simplest case of such situation is a particle with a spin. Although the fact that the wave function is multicomponent is not necessarily associated with a spin (an analogous situation takes place, for example, in the problems of atomic scattering [17–20]), we will restrict our consideration to this situation only. Thus the potential $\hat{\mathbf{A}}$ (the field $\hat{\mathbf{F}}$) will be named the *spin gauge potential* (*field*).

We now turn our attention to the consistent qualitative description of this gauge field. Assume that the particle with a spin is described by the wave function $\vec{\psi}$ of $n \geq 2$ components (we will denote multicomponent wave function with a arrow). In the case of the scalar (single-component) function $\psi$ considered in Subsection 1.1, it is natural to require the equations' invariance with respect to the gauge transformations of the unitary group $\mathbf{U}(1)$, which results in the introduction of the electromagnetic field. For the multicomponent function $\vec{\psi}$, it would appear reasonable to generalize this requirement by demanding the gauge invariance with respect to the transformations of the unitary group $\mathbf{U}(n)$. It is just these transformations that leave the absolute value of the wave function (the observable value) invariant. (Note that $\mathbf{U}(n) = \mathbf{U}(1) \times \mathbf{SU}(n)$ and the gauge potential in $\mathbf{U}(1)$ describes the evolution of a particle scalar characteristic, which is naturally associated with an *electric charge*, while the gauge potential in $\mathbf{SU}(n)$ describes the nontrivial dynamics associated with the evolution of the vector internal characteristic of a particle such as *spin* (see [45]).)

The group $\mathbf{U}(n)$ is non-Abelian, and hence the spin gauge potential $\hat{\mathbf{A}}$ is, in the general case, non-Abelian as well. As is known [4], under the gauge transformations, the non-Abelian gauge potential is transformed along with the wave function as

$$\hat{\mathbf{A}} \rightarrow \hat{U}^{-1} \hat{\mathbf{A}} \hat{U} + i \hat{U}^{-1} \frac{\partial \hat{U}}{\partial \boldsymbol{\mu}} \ , \quad \vec{\psi} \rightarrow \hat{U} \vec{\psi} \ , \tag{1.7}$$

where $\hat{U}(\boldsymbol{\mu}) \in \mathbf{U}(n)$ is a local unitary transformation. Since we have no physical evidences (observable data) for a particle with a spin, supporting the existence of a certain physical field corresponding to the group $\mathbf{U}(n)$, assume $\hat{\mathbf{A}}(\boldsymbol{\mu}) \equiv 0$ in the initial frame of reference, in which the equations are formulated. Then an arbitrary unitary transformation will induce a *pure gauge potential* (see Eq. (1.7))

$$\hat{\mathbf{A}} = i \hat{U}^{-1} \frac{\partial \hat{U}}{\partial \boldsymbol{\mu}} \ . \tag{1.8}$$

The zero field (curvature) corresponds to this potential:

$$\hat{F}_{ij} = \frac{\partial \hat{A}_j}{\partial \mu_i} - \frac{\partial \hat{A}_i}{\partial \mu_j} - i[\hat{A}_i, \hat{A}_j] \equiv 0 \ . \tag{1.9}$$

From here on square brackets denote a commutator. Therefore a pure gauge field, Eqs. (1.8), (1.9), has no force effect on the particle. However the non-Abelian gauge field possesses a number of features that distinguish it from the usual Abelian electromagnetic field. Specifically, for the non-Abelian field, its flux through a closed contour cannot be defined. Indeed, the divergence of the non-Abelian field is nonzero (see [5]), while the term in the integrand in the Stokes formula, which is similar to Eqs. (1.4), (1.6), is not a field tensor. (That is why formula (1.6) is not true beyond the limits of Abelian (adiabatic) approximation.) As is shown in [5], it is natural to use the phase contour integral $\oint \hat{\mathbf{A}} \, d\boldsymbol{\mu}$ instead of the flux for the non-Abelian field. This value is nonzero even for pure gauge potential (1.8) and, as will be seen, may give rise to observable physical effects. (This value is important in the theories of non-Abelian fields and is given in terms of the phase factor like $\mathrm{P} \exp(i \oint \hat{\mathbf{A}} \, d\boldsymbol{\mu})$, which is called the Wilson loop [92]. Here $\mathrm{P}$ is the operator of chronological ordering. This factor is transformed covariantly and may results in the observable physical effects even with zero field strength [5].)

Note that among all possible unitary transformations $\vec{\psi} \rightarrow \hat{U} \vec{\psi}$, some transformations are special in a certain sense. Namely, a unitary transformation (more precisely, a family of the unitary transformations) exists that brings the initial Hermitian Hamiltonian into the diagonal form: $\hat{U}^{-1} \hat{H} \hat{U} = \hat{H}_d$. This transformation implies the passage to a local basis associated with the Hamiltonian eigenvectors. At that the Hamiltonian acquires a diagonal form, which simplifies the solution of the equation, however non-diagonal pure gauge potential (1.8) is introduced. In essence, it describes a certain *inertia*, which arises in attempting to follow locally the basis of the Hamiltonian eigenvectors. Thus with diagonalization transformation like (1.7), (1.8) we actually *replace the nontrivial (non-diagonal) matrix structure of the initial Hamiltonian with non-diagonal induced gauge potential (1.8).*

D e f i n i t i o n  I:  Pure gauge potential (1.8) that corresponds to the local unitary transformation $\hat{U}(\boldsymbol{\mu})$ such that the Hamiltonian $\hat{U}^{-1}\hat{H}\hat{U}$ acquires a diagonal form will be called *the spin gauge potential*.

Thus we actually *consider a particle with a spin as a multiplet of independent scalar particles (various spin states) in the external spin gauge field (1.8), (1.9). This gauge field describes completely the nontrivial evolution of isolated states and the transitions between the states*.

Notice that spin gauge potential (1.8) is not defined uniquely. Indeed, there exist a family of unitary matrices $\hat{U}$ that diagonalize the Hermitian matrix. This family is associated with the ambiguity in the definition of a system of normalized eigenvectors of the Hermitian matrix. Each eigenvector is defined up to an arbitrary phase factor like $\exp(i\phi)$ (see Remark 1 below). As a result, the local transformation

$$\vec{\psi} \to \exp(i\hat{\phi})\vec{\psi} \quad \text{or} \quad \hat{U} \to \hat{U}\exp(i\hat{\phi}), \quad \text{where} \quad \hat{\phi} = \text{diag}(\phi_1,...,\phi_n) \quad (1.10)$$

does not change the diagonalizing properties of the matrix $\hat{U}$. Consequently, transformation (1.10) specifies the transformation of the local gauge invariance of a spin gauge potential. With transformations (1.10) it changes as

$$\hat{\mathbf{A}} \to e^{-i\hat{\phi}}\hat{\mathbf{A}}e^{i\hat{\phi}} - \frac{\partial\hat{\phi}}{\partial\boldsymbol{\mu}} . \quad (1.11)$$

It follows (compare with Eq. (1.2a)) that gauge transformations (1.10), (1.11) of spin gauge potential (1.8) correspond to the Abelian subgroup $\mathbf{U}^n(1) = \underbrace{\mathbf{U}(1)\times...\times\mathbf{U}(1)}_{n}$ of the initial group $\mathbf{U}(n)$. In particular, this implies that in the adiabatic approximation, where all the spin states are considered as independent ones, the action of the gauge group $\mathbf{U}^n(1)$ is accountable for $n$ independent $\mathbf{U}(1)$ gauge fields, which are similar to the electromagnetic field (see below).

R e m a r k  I:  In this paper we discuss a particle with nondegenerate levels, that is, with the distinct Hamiltonian eigenvalues (they may coincide only at the intersection points of the terms). Generally, when a level has a constant $k$-fold degeneracy, the corresponding gauge sector is given by the non-Abelian group $\mathbf{U}(k)$ (see [9,10,18,60,61,93]). This sector is conserved after reducing to the adiabatic approximation (see below).

Now let us describe the action of spin gauge field (1.8) on the particle evolution. So, we diagonalize the Hermitian Hamiltonian by introducing pure gauge non-Abelian potential (1.8), for which the flux-analogue is nonzero: $\oint \hat{\mathbf{A}}\,d\boldsymbol{\mu} \neq 0$. Note that under the gauge transformations $\mathbf{U}^n(1)$, Eqs. (1.10), (1.11), the gradient term $\partial\hat{\phi}/\partial\boldsymbol{\mu}$ does not contribute into this value. Hence the integrand is in fact transformed covariantly and the value $\oint \hat{\mathbf{A}}\,d\boldsymbol{\mu} \neq 0$ can be observable. This causes Berry phase (1.5) to occur, and consequently, we can observe the interference effects akin to the Aharonov-Bohm effect in an electromagnetic field. As mentioned above, the integral $\oint \hat{\mathbf{A}}\,d\boldsymbol{\mu}$ is determined by the orientation of the integration contour with respect to the degeneracy points of the Hamiltonian eigenvalues, which play the role of 'magnetic monopoles'. Structures of the 'magnetic monopole' type for non-Abelian fields and the non-Abelian analogues for the Aharonov-Bohm effect were first discussed in [5]. To all appearance, we are dealing here with exactly such effects. At the same time, both purely phase effects (Berry phase) and various transport effects, similar to Shelankov displacement [3], may arise. The latter correspond to various manifestations of the topological spin transport [17–91].

It is vital to note the fundamental difference between the system's geometry in the Berry case under consideration and in the case of the electromagnetic Aharonov-Bohm effect. Indeed, for the infinite thin solenoid we have a 1-dimensional singularity (a line) in a 3-dimensional space. Hence the geometry of the particle motion in the Aharonov-Bohm effect in fact is *2-dimensional* (one may disregard the coordinate directed along the solenoid). This, in particular, results in the one-to-one relation between Dirac phase (1.4a) and the flux in the solenoid, and consequently, the Dirac quantization conditions follow. As to the case of the degeneracy points of the Hermitian Hamiltonian parameter space, the singularities have the dimensionality of zero (points), and the geometry of the particle evolution is *3-dimensional*. This gives the additional degree of freedom, and as a consequence, Berry phase (1.6) is determined not only by the 'topological charge' of the singularity, but also by the *spatial orientation of the contour* with respect to this singularity.

In summary, based on the 'phase' formalism developed in [5] for gauge fields, by analogy with the statement quoted at the end of Subsection 1.1, we can say: *the topological spin effects are the gauge-invariant manifestation of the nonintegrable Berry phase factor*. (Here, gauge invariance (1.10), (1.11) with respect to the group $\mathbf{U}^n(1)$ is implied.) Let's stress once again that the introduction of the pure gauge potential (1.8) together with the attendant effects is nothing but a method to describe the solution to the initial equation, which is free of any physical field $\hat{\mathbf{A}}$. However, for a number of problems, this way turns out to be a very convenient one, or even the only one possible. Indeed, for an equation with the matrix operator its diagonalization is exactly the same as its solution.

**1.4. A reduction to the adiabatic approximation.** The action of the spin gauge potential (1.8) is most descriptive in the *adiabatic* evolution approximation. In the adiabatic case the Hamiltonian $\hat{H}$ and the diagonalizing transformation $\hat{U}$ are smoothly dependent upon the parameters, while the derivatives of these values are small (of the order of the adiabaticity parameter $\varepsilon$). Consequently, the potential (1.8) is also small in magnitude. This gives grounds to neglect its off-diagonal elements (see [94,95]), which corresponds to neglecting the transitions between different current particle eigenstates. Then potential (1.8) is reduced to the diagonal adiabatic potential

$$\hat{\mathbf{A}}^{(ad)} = \mathrm{dg}\,\hat{\mathbf{A}} = \mathrm{dg}\!\left(i\hat{U}^{-1}\frac{\partial \hat{U}}{\partial \boldsymbol{\mu}}\right). \qquad (1.12)$$

Here and further the operator dg makes the off-diagonal matrix elements vanish. With the reduction (1.12), the potential keeps its properties and the phase form $\oint \hat{\mathbf{A}}^{(ad)} d\boldsymbol{\mu}$ is nonzero as before. But the diagonal potential (1.12) becomes the *Abelian* one (the diagonal matrices commutate), which affects substantially the gauge field tensor $\hat{\mathbf{F}}$. Indeed, in the Abelian case, commutator in Eq. (1.9) vanishes, and the expression for the tensor $\hat{\mathbf{F}}^{(ad)}$ takes the form of Eq. (1.1):

$$\hat{F}_{ij}^{(ad)} = \frac{\partial \hat{A}_j^{(ad)}}{\partial \mu_i} - \frac{\partial \hat{A}_i^{(ad)}}{\partial \mu_j} \neq 0 \,. \qquad (1.13)$$

D e f i n i t i o n  II: The Abelian potential (1.12), which is composed of the diagonal elements of the spin gauge potential (1.8), and the associated force field (1.13) will be referred to as the *adiabatic spin gauge potential* and the *adiabatic spin gauge field*, respectively. (Alternative names are the '*Berry gauge potential*' and '*Berry gauge field*'.)

Thus, when passing on to adiabatic approximation (1.12), the intensity of the gauge field (curvature) (1.9), (1.13) becomes nonzero, while the field is found to be not purely gauge but the *force* one. The rejection of the off-diagonal elements of the gauge potential substantially affects the field tensor, because the filed is a *nonlinear* function of the potential in the non-Abelian case, Eq. (1.9). It is easy to realize from the symmetry considerations why in the adiabatic approximation we arrive at a new nontrivial field. The point is that in the adiabatic approximation a dynamic system acquires $n$ new invariants: *adiabatic invariants of action*.

Accordingly, the system acquires *n* new symmetries: the *angle* variable, which is conjugate to the canonical action variable, turns out to by cyclic. It is the local gauge invariance of the equations with respect to shifts of *n* angle variables that gives rise to *n* gauge fields similar to electromagnetic ones (in Eqs. (1.12), (1.13) all these fields are combined into one diagonal matrix).

Within the limits of the adiabatic approximation, we have passed on to nontrivial field (1.13), which can make a force action on the particle motion. This field is diagonal, and hence, the whole problem is diagonalized. In view of this fact, the evolution of separate components (eigenstates) can be treated independently. As a result, we have *n* nontrivial fields (components of Eq. (1.13)), each acting on its corresponding state of the particle and possessing the gauge symmetry $U(1)$ (see Sections 2 and 3 below). These fields are identical with the electromagnetic fields and, consequently, must result in the similar force terms in the motion equations. Various cases for the force influence of this field have been discussed in papers on the topological spin transport [17–91], while the general form of the theory is presented in this work.

The transition from the exact forceless field (1.8), (1.9) with the Aharonov-Bohm-type interference effects to the force adiabatic field (1.12), (1.13) is the result of the approximate solution description in the adiabatic limit. In other words, *the transition to the adiabatic case replaces a purely phase action of exact forceless field (1.8), (1.9) with the approximate efficient force action of field (1.12), (1.13)*. Actually both of these descriptions have to give identical results within the accuracy of the adiabatic approximation.

**1.5. About the present paper.** In this work we present a general theory of the motion of a nonrelativistic quantum particle with a spin in external fields including spin gauge field. Special attention is given to the adiabatic case, where the field $\hat{\mathbf{F}}^{(ad)}$ is nonzero and its force action manifests itself in the motion equations. We show that in the general case spin gauge fields act over the generalized phase space of a particle, i.e., in the coordinate and momentum spaces and in the time. The theory is reduced into convenient 4-dimensional form, where the adiabatic spin gauge field is seen to be an analogue to *n* electromagnetic fields. The only difference is that this field acts in an 8-dimensional phase space of 4-coordinates and 4-momentums.

After presenting the general theory, we consider a number of concrete examples (both known and new), namely, a spin 1/2 motion in the inhomogeneous and variable magnetic field, a spin 1/2 motion in electric and magnetic fields with consideration of the spin-orbit interactions, and electromagnetic waves (photons) propagation in inhomogeneous dielectric media within the limits of the geometrical optics approximation.

## 2. NONRELATIVISTIC THEORY

**2.1. Diagonalization of the Hamiltonian and spin gauge potential.** Let us consider a nonrelativistic particle with a spin in external fields, which is given by an $n \times n$ Hamiltonian-matrix. In the general case we have:

$$\hat{H} = \hat{H}(\mathbf{p}, \mathbf{r}, t) , \qquad (2.1a)$$

where **p** and **r** are the momentum and the coordinate operators, respectively, while *t* stands for the time. (We will write the differential operators without hats). (The explicit dependence on time or other variables in Eq. (2.1a) can be represented, as the convenience requires, in terms of some set of parameters, as in the theory of Berry phase; the conversions between different representations are exemplified below (Subsections 2.5 and 5.1, 5.2).)

A particle motion, as in classical mechanics, can be formally presented on the extended phase space

$$\mathbf{m} = (\mathbf{p}, \mathbf{r}, t) . \qquad (2.2)$$

Since variables **p** and **r** are non-commuting, one can question the possibility of diagonalization procedure on **m**-space and of introduction of the spin gauge potential. However, in two important particular cases such approach is valid. Firstly, this is a frequently considered in the literature case, when the non-diagonal part of the Hamiltonian depends either on **p** only [47,54,56–75,77–84,86,89–91] or on **r** and $t$ (see [17–53]). Secondly, this is the semiclassical approximation, when a freely propagating particle takes the form of a wave packet, and one can neglect commutators **p** and **r** ([55,76], see also Subsection 2.3). We assume further that we deal with one of the above-mentioned cases, and leave the issue of a non-diagonal Hamiltonian on the space of non-commuting variables open. In terms of Eq. (2.2), Hamiltonian (2.1a) has the form

$$\hat{H} = \hat{H}(\mathbf{m}) . \qquad (2.1)$$

An important particular case of Eq. (2.1) is the Hamiltonian linear in the spin operator $\hat{\boldsymbol{\sigma}}$:

$$\hat{H} = \hat{I} H_0(\mathbf{m}) + \hbar \hat{\boldsymbol{\sigma}} \mathbf{H}_1(\mathbf{m}) . \qquad (2.3)$$

Here $\hat{I}$ is the unit operator in the spin space. Since $\hat{\boldsymbol{\sigma}}$ is the operator-vector, we should take a scalar product of this operator with some other vector $\mathbf{H}_1$, which is given in Eq. (2.3). In the paper we will consider a general theory for Hamiltonian (2.1) with simple levels along with the examples of how to apply this theory to the Hamiltonians like (2.3).

Hamiltonian (2.1) describes an $n$-level system and, in a general case, has a non-diagonal form. Deducing the independent spin states of the system is equivalent to the Hamiltonian diagonalization. To diagonalize Hamiltonian (2.1), let us make the transformation of variables $\vec{\psi} \to \hat{U}\vec{\psi}$, which is defined by the unitary matrix-operator $\hat{U}(\mathbf{m})$. With the similarity transformation the Hamiltonian becomes diagonal:

$$\hat{H}_d = \hat{U}^{-1}\hat{H}\hat{U} = \mathrm{diag}(E_1,...,E_n) , \qquad (2.4)$$

where $E_i = E_i(\mathbf{m})$ are the current eigenvalues of the initial Hamiltonian $\hat{H}$ [112]. We assume that the particle does not meet the degeneracies during its evolution:

$$E_i \neq E_j \text{ for } i \neq j . \qquad (2.5)$$

Under diagonalizing transformation (2.4) on the **m**-space, a pure gauge potential (connection) occurs (see Subsection 1.3 and Eq. (1.8))

$$\hat{\mathbf{A}}(\mathbf{m}) = i\hat{U}^{-1}\frac{\partial \hat{U}}{\partial \mathbf{m}} . \qquad (2.6)$$

This is the *spin gauge potential* $\hat{\mathbf{A}}$ on the extended phase **m**-space. Obviously, it can be written in the space components

$$\hat{\mathbf{A}} = (\hat{\mathbf{A}}_\mathbf{p}, \hat{\mathbf{A}}_\mathbf{r}, \hat{A}_t) , \qquad (2.7)$$

where $\hat{\mathbf{A}}_\mathbf{p} = i\hat{U}^{-1}\partial\hat{U}/\partial\mathbf{p}$, etc. Due to the action of the gauge potential (2.6), (2.7), we should replace the conventional derivatives with the *covariant* ones, namely:

$$i\frac{\partial}{\partial \mathbf{m}} \to i\frac{\partial}{\partial \mathbf{m}} + \hat{\mathbf{A}} \qquad (2.8)$$

or

$$i\frac{\partial}{\partial \mathbf{p}} \to i\frac{\partial}{\partial \mathbf{p}} + \hat{\mathbf{A}}_\mathbf{p} , \quad i\frac{\partial}{\partial \mathbf{r}} \to i\frac{\partial}{\partial \mathbf{r}} + \hat{\mathbf{A}}_\mathbf{r} , \quad i\frac{\partial}{\partial t} \to i\frac{\partial}{\partial t} + \hat{A}_t . \qquad (2.8a)$$

Since in the corresponding representations we have

$$\mathbf{p} = -i\hbar\frac{\partial}{\partial \mathbf{r}} , \quad \mathbf{r} = i\hbar\frac{\partial}{\partial \mathbf{p}} , \quad \hat{H} = i\hbar\frac{\partial}{\partial t} \qquad (2.9)$$

(the last formula is the Schrödinger equation), the replacement of derivatives with covariant derivatives (2.8a) implies a shift of all dynamic variables in the Hamiltonian, or in other words, a transition to the *generalized* values.

As a result of the transition to the generalized dynamic variables under the action of gauge potential (2.6), (2.7), the diagonalized Hamiltonian (2.4) should be written as

$$\hat{H} = \hat{H}_d\left(\mathbf{P} - \hbar\hat{\mathbf{A}}_r, \mathbf{R} + \hbar\hat{\mathbf{A}}_p, t\right) - \hbar\hat{A}_t . \qquad (2.10)$$

Here

$$\mathbf{P} = \mathbf{p} + \hbar\hat{\mathbf{A}}_r , \quad \mathbf{R} = \mathbf{r} - \hbar\hat{\mathbf{A}}_p \qquad (2.11)$$

are the generalized (or canonical) momentum and coordinates, respectively, whereas $\mathbf{p}$ and $\mathbf{r}$ are the ordinary (covariant) ones. Hamiltonian (2.10) is exact and represents the particle evolution in the new coordinates. It is still non-diagonal; however, this non-diagonality is now entirely contained in the spin gauge potential (2.6). It is also seen from Eq. (2.10) that the spin gauge potentials $\hat{\mathbf{A}}_p$ and $\hat{\mathbf{A}}_r$ act in much the same way as the vector-potentials of the electromagnetic field do in the momentum and the coordinate spaces. At the same time, the temporal term $\hat{A}_t$ can be presented as an analogue of the scalar potential of an electromagnetic field (see Subsections 3.2 and 3.3 below). From equations (2.4), (2.10), and (2.11) a general expression for particle's Berry phase also follows, which will be derived in Section 3 in the more convenient 4-dimensional form.

**2.2. Spin gauge field tensor and transition to the adiabatic case.** Owing to its non-diagonality, the spin gauge potential $\hat{\mathbf{A}}$ (2.6) is, in general, non-Abelian. Let us consider a tensor of the spin gauge field (curvature) $\hat{\mathbf{F}}(\mathbf{m})$ associated with the spin gauge potential $\hat{\mathbf{A}}(\mathbf{m})$ over the $\mathbf{m}$-space:

$$\hat{F}_{ij} = \frac{\partial \hat{A}_j}{\partial m_i} - \frac{\partial \hat{A}_i}{\partial m_j} - i\left[\hat{A}_i, \hat{A}_j\right] \equiv 0 . \qquad (2.12)$$

It is easy to verify through direct calculations, by substituting Eq. (2.6) into Eq. (2.12), that the tensor $\hat{\mathbf{F}}$ equals zero (see Subsection 1.3). Hence the field $\hat{\mathbf{F}}$ is *forceless*. At the same time, the contour integral $\oint \hat{\mathbf{A}} d\mathbf{m}$ is, generally speaking, nonzero. As mentioned in Subsections 1.2 and 1.3, its value can be associated with the presence of the degeneracies (term intersections) in the $\mathbf{m}$-space, and depends on the orientation of the integration contour with respect to these points. (It should be noted that in the $\mathbf{m}$-space the degeneracies have the form of 4-dimensional hypersurfaces, since this space is 7-dimensional in our case. However, they can be considered as points in other convenient parameter space or in a subspace of the $\mathbf{m}$-space, see Subsection 2.5 and Section 5.)

The off-diagonal terms of the gauge potential $\hat{\mathbf{A}}$ are responsible for transitions between different spin levels of the particle. If we consider the adiabatic evolution of the particle, the off-diagonal terms can be neglected. They make a contribution of the second order in the adiabaticity parameter $\varepsilon$ (see next Subsection 2.3) [94,95]. By rejecting the off-diagonal terms of $\hat{\mathbf{A}}$, we arrive at the Abelian potential:

$$\hat{\mathbf{A}}^{(ad)} = \mathrm{dg}\,\hat{\mathbf{A}} = i\,\mathrm{dg}\left(\hat{U}^{-1}\frac{\partial \hat{U}}{\partial \mathbf{m}}\right) . \qquad (2.13)$$

Then, the expression for the intensity field tensor is given by the formula

$$\hat{F}_{ij}^{(ad)} = \frac{\partial \hat{A}_j^{(ad)}}{\partial m_i} - \frac{\partial \hat{A}_i^{(ad)}}{\partial m_j} = i\,\mathrm{dg}\left(\frac{\partial \hat{U}^{-1}}{\partial m_i}\frac{\partial \hat{U}}{\partial m_j} - \frac{\partial \hat{U}^{-1}}{\partial m_j}\frac{\partial \hat{U}}{\partial m_i}\right) \neq 0 . \qquad (2.14)$$

The spin gauge field tensor is now nonvanishing, and the field $\hat{\mathbf{F}}^{(ad)}$ is a *force* one (see Subsection 1.4). The degeneracies in the spectrum of the initial Hamiltonian are the points of singularity of $\hat{\mathbf{F}}^{(ad)}$. In the generic case, the singularity of $\hat{\mathbf{F}}^{(ad)}$ is of the form of the Dirac monopole in the corresponding 3-dimensional space, while the integral $\oint \hat{\mathbf{F}}^{(ad)} d\mathbf{s}$ over the surface enveloping the singularity is equal to the *topological charge* (the Chern number) of this monopole [7–9]. (As a consequence, when the surface intersects the singularity, the integral $\int \hat{\mathbf{F}}^{(ad)} d\mathbf{s}$ abruptly changes its value, which corresponds, for example, to the quantum Hall effect, see in [7,8].)

The tensor of the adiabatic spin gauge field (2.14) is an invariant with respect to the following gauge transformations of the potential:

$$\hat{\mathbf{A}}^{(ad)} \to \hat{\mathbf{A}}^{(ad)} - \frac{\partial \hat{\phi}}{\partial \mathbf{m}} , \qquad (2.15)$$

where $\hat{\phi}$ stands for an arbitrary diagonal matrix. This corresponds to gauge invariance (1.10), (1.11) with respect to the $\mathbf{U}^n(1)$ group affecting independently the independent spin states of the particle.

**2.3. About adiabatic approximation.** By an adiabatic approximation we mean the absence of transitions between different terms in the course of the system's evolution. Let us consider prerequisites to the absence of such transitions in various cases.

First we take the *localized* particle states with a discrete spectrum. In this case, the $\mathbf{r}$-dependence of the Hamiltonian (2.1) only determines the discrete spectrum of the system (the levels), while the $t$-dependence may cause the transitions between different stationary states. Then the adiabatic approximation is determined by the slowness of the time-dependence of the Hamiltonian. The relevant small parameter is

$$\varepsilon = \frac{\hbar}{\delta E^2}\left|\frac{\partial E}{\partial t}\right| \ll 1 , \qquad (2.16)$$

where $E = E_k$ is the current energy value and $\delta E = E_k - E_{k-1}$ is the current distance to the adjacent spin level. The fact that the denominator of Eq. (2.16) has to contain the squared distance to the neighboring term is demonstrated, for example, in [96].

Now let's turn our attention to *free (nonlocalized)* motion of the particle with the continuous energy spectrum. In this case, the functions $E(\mathbf{p})$ determine the particle terms (dispersion curves) and the transitions between the terms can result from the time inhomogeneity of the Hamiltonian as well as from its spatial inhomogeneity. In the former case, the transitions between the energy levels occur with the constant momentum, whereas in the latter case, we have the transitions between different momentum states with the fixed energy. Hence the small parameter specifying the adiabatic approximation in terms of the absence of the transitions between the terms of different spin states is

$$\varepsilon = \hbar \max\left(\frac{1}{\delta E^2}\left|\frac{\partial E}{\partial t}\right| , \frac{1}{\delta p^2}\left|\frac{\partial p_i}{\partial r_j}\right|\right) \ll 1 . \qquad (2.17)$$

Here $E$ and $\delta E$ are the particle energy and the distance to the adjacent spin term for given $(\mathbf{p}, \mathbf{r}, t)$, while $\mathbf{p}$ and $\delta \mathbf{p}$ are the particle momentum with the fixed $(E, \mathbf{r}, t)$ and the distance to the nearest spin term (see also equation (3.15) below). (The derivatives $\partial p_i / \partial r_j$ describe the variations in the particle momentum with a given energy as a consequence of the Hamiltonian spatial inhomogeneity.) The first term in Eq. (2.17) describes the adiabaticity parameter (2.16) associated with the time-dependence of the Hamiltonian, whereas the second term determines the absence of the term transitions caused by the scattering on spatial inhomogeneities. Note that the second term in Eq. (2.17) ensures the validity of the *semiclassical* approximation. It is precisely this fact that allows considering a free motion of the particle with the simultaneously determined (from the standpoint of the wave packet motion) momentums and coordinates, as formula (2.17) implies. From here on we will think of adiabaticity in a general sense Eq. (2.17) for the freely propagating particles.

The above-analyzed cases are associated with the simplest situation, where the system possesses just two characteristic spatial and temporal scales, namely, the de Broglie scales of the particle and characteristic scales of the Hamiltonian inhomogeneities. It is evident that more complex, intermediate cases with many scales are also possible; sometimes they can be reduces to one of the two simple models. Two examples can be offered.

Firstly, an additional scale exists when a particle is moving in a smoothly-inhomogeneous magnetic field – this is the Larmor radius of the given state of the particle. If the typical scale of the field inhomogeneity is much larger than the Larmor radius, we can average the motion over the scale of this Larmor radius. As a result, we reduce the problem to the problem of a particle motion within the scales much larger than the Larmor radius. Then, criterion (2.17) should be invoked, where the value $\left|\partial p_i / \partial r_j\right|$ is also averaged over the scale of the Larmor radius and, as a consequence, yields a large scale of the inhomogeneity.

Secondly, another example is the quasi-stationary states in the double-humped potential (see, for example, [97]). In this situation, the characteristic state lifetime is an additional temporal scale. If the lifetime is large as compared to other typical times of the problem, its quasi-stationary states can be treated within the limits of the approximation of the discrete spectrum and the finite trajectories, and hence, we can invoke adiabaticity criterion (2.16). Notice that the localized particle's states of this kind have the topological contribution in the tunnel transport through the potential barriers adiabatically varying in time (see, for example, [48,50]).

Note finally that the use of the diagonalized Hamiltonian (2.10) with the Abelian potentials (2.13) implies neglecting of the terms of the order of $\varepsilon^2$ and higher. However, it is well known that nonadiabatic transitions between nondegenerate terms are of order of $\exp(-1/\varepsilon)$ (see, for example, [98]). It has been shown for a number of cases that the diagonalization procedure in question can be continued by the iterative technique, which is the continuation of the first approximation in $\varepsilon$ as an asymptotic series with the reminder term of the order of $\exp(-1/\varepsilon)$ [95,99].

**2.4. Commutation relations and equations of motion.** Let's consider how the presence of the spin gauge potential (2.6)–(2.10) affects the commutation relations between the dynamic variables and the equations of motion. In the exact case $\hat{\mathbf{F}} \equiv 0$, Eq. (2.12), these relations are standard, however in the adiabatic approximation $\hat{\mathbf{F}}^{(ad)} \neq 0$, Eq. (2.14), and we have from Eqs. (2.8) and (2.9) that (compare with [86])

$$\left[p_i, p_j\right] = i\hbar^2 \hat{F}^{(ad)}_{r_i r_j} \, , \quad \left[r_i, r_j\right] = i\hbar^2 \hat{F}^{(ad)}_{p_i p_j} \, , \quad \left[r_i, p_j\right] = i\hbar \delta_{ij} \hat{I} - i\hbar^2 \hat{F}^{(ad)}_{p_i r_j} \, . \qquad (2.18)$$

Here we introduce components of the gauge field $\hat{\mathbf{F}}^{(ad)}$ as (compare with (2.14) and [55,76])

$$\hat{F}^{(ad)}_{r_i r_j} = \frac{\partial \hat{A}^{(ad)}_{r_j}}{\partial r_i} - \frac{\partial \hat{A}^{(ad)}_{r_i}}{\partial r_j} = i \, \mathrm{dg}\left(\frac{\partial \hat{U}^{-1}}{\partial r_i}\frac{\partial \hat{U}}{\partial r_j} - \frac{\partial \hat{U}^{-1}}{\partial r_j}\frac{\partial \hat{U}}{\partial r_i}\right),$$

$$\hat{F}^{(ad)}_{p_i r_j} = \frac{\partial \hat{A}^{(ad)}_{r_j}}{\partial p_i} - \frac{\partial \hat{A}^{(ad)}_{p_i}}{\partial r_j} = i \, \mathrm{dg}\left(\frac{\partial \hat{U}^{-1}}{\partial p_i}\frac{\partial \hat{U}}{\partial r_j} - \frac{\partial \hat{U}^{-1}}{\partial r_j}\frac{\partial \hat{U}}{\partial p_i}\right),$$

etc. Commutation relations (2.18) demonstrate the analogy with an electromagnetic field.

The Hamiltonian motion equations have the usual canonical form in the generalized variables:

$$\dot{\mathbf{P}} = -\frac{i}{\hbar}\left[\mathbf{P}, \hat{H}\right], \quad \dot{\mathbf{R}} = -\frac{i}{\hbar}\left[\mathbf{R}, \hat{H}\right], \qquad (2.19)$$

where the dot stands for the full derivative with respect to time, and Hamiltonian (2.10) is taken as a function of generalized variables: $\hat{H} = \hat{H}(\mathbf{P}, \mathbf{R}, t)$. The momentum and coordinates that are measured by an observer are, in fact, covariant quantities $\mathbf{p}$ and $\mathbf{r}$. The motion equations for these have a canonical form as well:

$$\dot{\mathbf{p}} = -\frac{i}{\hbar}\left[\mathbf{p}, \hat{H}\right], \quad \dot{\mathbf{r}} = -\frac{i}{\hbar}\left[\mathbf{r}, \hat{H}\right], \qquad (2.20)$$

however here Hamiltonian (2.10) should be considered as a function of variables **p** and **r**: $\hat{H} = \hat{H}(\mathbf{p}, \mathbf{r}, t)$. Non-trivial commutation relations (2.18) of these variables in adiabatic approximation lead to the appearance of the additional terms, connected to the spin gauge field action. Indeed, in the adiabatic and semiclassical approximations we have from (2.10), (2.18) and (2.20) the following motion equations:

$$\dot{\mathbf{p}} = -\frac{\partial \hat{H}_d}{\partial \mathbf{r}} + \hbar \hat{\mathbf{F}}^{(ad)}_{\mathbf{rm}} \dot{\mathbf{m}} \; , \quad \dot{\mathbf{r}} = \frac{\partial \hat{H}_d}{\partial \mathbf{p}} - \hbar \hat{\mathbf{F}}^{(ad)}_{\mathbf{pm}} \dot{\mathbf{m}} \; . \tag{2.21}$$

Here the convolution of the second subscripts of the field tensors with the following vectors is assumed.

Equations (2.21) are the basic motion equations of an adiabatic semiclassical particle, which take into account the force action of the spin gauge field $\hat{\mathbf{F}}^{(ad)}$. Similar equations have been derived in [55] in the course of the Lagrange description of the wave packet dynamics. Particular cases of these are discussed in many papers on the topological spin transport [21–26,54–59,63,65,68,76,84,86–91]. The first terms in the right-hand sides of equations (2.21) represent the evolution specified by the current eigenvalues of the initial Hamiltonian (see Eq. (2.4)). The second terms in the right-hand sides of equations (2.21) constitute the corrections that describe *topological spin transport* of the particle. These terms are proportional to $\hat{\mathbf{F}}^{(ad)}$.

As a consequence of the adiabatic approximation diagonality, equations (2.18) and (2.21) decompose into sets of $n$ independent equations, which describe the adiabatic evolution of the particle's independent spin states. The gauge potential enters into the equations of motion only through the field intensity tensor, and hence, the equations of motion are gauge-invariant with respect to the group $\mathbf{U}^n(1)$ (2.15). Finally, one can readily see (see also Subsection 3.3) the analogy between the force action of the adiabatic spin gauge field and the force action of the electromagnetic field. The terms related to **p** and **r** components of $\hat{\mathbf{F}}^{(ad)}$ in Eqs. (2.21) present the analogues of the Lorentz force in the coordinate and the momentum spaces, while the terms involving $t$-components of $\hat{\mathbf{F}}^{(ad)}$ present the analogues of the electric field action.

R e m a r k  II: The spin analogues of the electric field, which are described in the present paper, have no relations to 'geometric electric field' that is considered, for example, in [20–23]. Indeed, the latter is of order of $\hbar^2$ (rather than $\hbar$) and is connected to the deviation of the classical magnetic moment from the magnetic field direction. In quantum adiabatic case this 'geometric electric field' field is equal to zero for pure states of spin 1/2. An analogue of the electric field, which corresponds to the one introduced in our paper, is presented in the paper [26] that appeared at the time when the present paper was being prepared for publication.

The forces connected with tensor of the adiabatic spin gauge field in Eqs. (2.21) are proportional to the projection of the particle spin (see Eqs. (2.22)–(2.24) below) and are not in a direct relationship with its charge. Hence *different spin states of the particle give rise to essentially distinct spin charges of the particle in the spin gauge field* $\hat{\mathbf{F}}^{(ad)}$, i.e. the particle can be represented as a *spin multiplet*. For example, an electron in a magnetic field possesses two states with the spin charge +1/2 and −1/2 (in units of $\hbar$).

When a particle is influenced by an external electromagnetic field, the electromagnetic potentials are added into equations (2.10) and (2.11), while in the first equations of (2.18) and (2.21) the following terms arise [113]:

$$[p_i, p_j] = i\hbar \frac{e}{c} \varepsilon_{ijk} \mathcal{B}_k + i\hbar^2 \hat{F}^{(ad)}_{r_i r_j} \; , \tag{2.18a}$$

$$\dot{\mathbf{p}} = -\frac{\partial \hat{H}_d}{\partial \mathbf{r}} + e\mathcal{E} + \frac{e}{c}(\dot{\mathbf{r}} \times \mathcal{B}) + \hbar \hat{\mathbf{F}}^{(ad)}_{\mathbf{rm}} \dot{\mathbf{m}} \; , \tag{2.21a}$$

where $\mathcal{E}$ and $\mathcal{B}$ are the electric and magnetic field, respectively, and $\varepsilon_{ijk}$ is the unit antisymmetric tensor. The analogy between the terms associated with the electromagnetic field and the spin terms will be clear in the four-dimensional form (Subsection 3.3).

In conclusion of this Subsection, let us determine qualitatively the forces that have a dominant role in particle evolutions of different types. First we consider a localized quasistationary state of the particle. Then it is evident that the particle's transport is weakly affected by the spin analogues of the magnetic field (since the particle is at rest). At the same time, *the spin analogue of the electric field will result in the slow transport of the localized states* (this is an adiabatic pumping under slow variations of potential barriers in time [48,50]). Consider now the opposite case of the freely moving particle. Clearly the basic variations in the particle's transport are associated with the forces responsible for the shift of the particle across the principal motion (the longitudinal shifts are small, being of order of $\varepsilon$ in the adiabatic approximation, and are not detectable on the background of the basic transport). Among them are primarily the spin analogues of the Lorentz force in the $\mathbf{p}$- and $\mathbf{r}$-spaces (see [17–26,39,47,54–63,65,68,76,84,86–91]). This is the reason why *the topological spin transport of free particles is frequently associated with the analogues of the magnetic field* and *the Hall effects*. Thus is, for example, the nature of the anomalous and intrinsic spin Hall effects [56–84] and the optical Magnus effect [86–91].

**2.5. Calculation of the tensor of the adiabatic spin gauge field.** The above-derived equations have been written for an arbitrary tensor of the spin gauge field $\hat{\mathbf{F}}^{(ad)}$. However, it turns out that it is frequently reducible to a certain universal form. As noted in Subsection 1.2, [6–10], the occurrence of the nonzero curvature tensor $\hat{\mathbf{F}}^{(ad)}$ is associated with the presence of degeneracy points (intersections of the spin terms) in the particle's spectrum. Let us consider the case of Hamiltonian (2.3), for which the spin operator contains the entire matrix structure.

Evidently, point $\mathbf{H}_1 = 0$, at which the spin term in the Hamiltonian vanishes, is the intersection point of $n$ spin terms. The vector $\mathbf{H}_1$ is a three-dimensional vector, while the field $\hat{\mathbf{F}}^{(ad)}(\mathbf{H}_1)$ in the $\mathbf{H}_1$-space takes the universal form of the magnetic monopole type [6–10]:

$$\hat{F}_{ij}^{(ad)}(\mathbf{H}_1) = -S\varepsilon_{ijk}\frac{H_{1k}}{H_1^3}, \text{ or } \mathbf{F}^{(ad)}(\mathbf{H}_1) = -S\frac{\mathbf{H}_1}{H_1^3}. \quad (2.22)$$

Here $S$ is the spin projection on the vector $\mathbf{H}_1$; it takes $n$ integer or half-integer values at the different spin states. In consequence of three-dimensionality of the $\mathbf{H}_1$-space, the antisymmetric tensor $\hat{\mathbf{F}}^{(ad)}$ can be represented in the form of pseudovector (2.22). (We will denote it by the same letter as the tensor, hoping that this will not lead to confusion.) Formula (2.22) demonstrates that the intersection point of the terms acts in the adiabatic approximation much as the Dirac magnetic monopole does. At the same time, the different values of the spin projection $S$ correspond to the different charges of the particle in the field generated by this monopole, resulting in the spin transport and the *spin splitting* of the particles.

From the equation (2.22), the value of the field $\hat{\mathbf{F}}^{(ad)}$ tensor can be calculated in any other convenient space. By using definitions (2.13) and (2.14) one can easily determine the rule for the field tensor transition from some $\mathbf{b}$-space into another $\mathbf{a}$-space:

$$\hat{F}_{ij}^{(ad)}(\mathbf{a}) = \frac{\partial b_k}{\partial a_i}\frac{\partial b_m}{\partial a_j}\hat{F}_{km}^{(ad)}(\mathbf{b}). \quad (2.23)$$

In particular, if $\mathbf{b} = \mathbf{H}_1$, while $\mathbf{a} = \mathbf{m}$, we have from Eqs. (2.22) and (2.23):

$$\hat{F}_{ij}^{(ad)}(\mathbf{m}) = -S\frac{\mathbf{H}_1}{H_1^3}\left(\frac{\partial \mathbf{H}_1}{\partial m_i} \times \frac{\partial \mathbf{H}_1}{\partial m_j}\right). \quad (2.24)$$

The above formula is the general one for the tensor of the adiabatic spin gauge field for the Hamiltonian of form Eq. (2.3). From Eq. (2.24) different components of $\hat{\mathbf{F}}^{(ad)}$ can be easily written in the components of the **m**-space (2.2). Various examples illustrating the application of the general formulas presented above will be given in Section 5.

## 3. FOUR-DIMENSIONAL FORM OF THE THEORY

**3.1. Hamiltonian formalism.** Relativistic equations of particle's motion and the spin gauge fields are outside the scope of this paper. However, owing to the analogy with the electromagnetic field, the basic notions of the above-suggested theory can be easily brough to the 4-dimentional form. This substantially simplifies the whole formalism and reveals the analogy with the electromagnetic field. The naturalness of this 4-dimensional form is a step towards the development of the relativistic theory.

Let us recall that the basic 4-vectors are the 4-vector of coordinates $r^\alpha = (ct, x, y, z)$ and the 4-vector of the momentum-energy $p^\alpha = (c^{-1}E, p_x, p_y, p_z)$. The Hamiltonian equations are representable in the 4-dimensional form. Since the Hamiltonian now should depend on energy as on an independent value, we will consider, instead of the usual Hamiltonian, the following one

$$\widetilde{H}(p^\alpha, r^\alpha) = H(\mathbf{p}, \mathbf{r}, t) - E . \tag{3.1}$$

Evidently this Hamiltonian vanishes on the wave function, which corresponds to the 4-dimensional Hamiltonian approach with couplings [100] as well as to the Hamiltonian occurring in the geometrical optics approximation [101]. The substitution of the quantum-mechanical operators in the form

$$p^\alpha = -i\hbar \frac{\partial}{\partial r_\alpha} , \tag{3.2}$$

results immediately in the following form of the Schrödinger equation:

$$|\widetilde{H}|\psi\rangle = 0 . \tag{3.3}$$

By using the Hamiltonian (3.1), the 4-dimensional Hamiltonian equations can be written in the usual form

$$\dot{p}^\alpha = -\frac{i}{\hbar}\left[p^\alpha, \widetilde{H}\right] , \quad \dot{r}^\alpha = -\frac{i}{\hbar}\left[r^\alpha, \widetilde{H}\right] . \tag{3.4}$$

In this case three basic components of equations (3.4) correspond to the ordinary Hamilton equations, while the zero component is associated with the equations

$$\dot{E} = \frac{\partial \widetilde{H}}{\partial t} , \quad \dot{t} = 1 , \tag{3.5}$$

which are clearly true.

**3.2. A spin gauge 4-potential and field tensor.** So, the Hamiltonian approach admits a rather concise 4-dimensional generalization. Let us consider now how the spin gauge potential and field can be represented in the framework of this approach. Let us introduce the 8-dimensional phase space of momentums and coordinates: $m^\alpha = (p^\alpha, r^\alpha)$. By analogy with the above derivations, we diagonalize the Hamiltonian (3.1). The energy added into Eq. (3.1) is a scalar in the spin space and in no way affects the performed transformations and the reasoning. Consequently, analogously to Eq. (2.6), the following gauge 4-potential is induced under the diagonalization:

$$\hat{A}^\alpha = i\hat{U}^{-1}\frac{\partial \hat{U}}{\partial m_\alpha} . \tag{3.6}$$

As in Eq. (2.7), this potential can be decomposed into the components of $m^\alpha$-space:

$$\hat{A}^\alpha = \left(\hat{A}_{p_\alpha}, \hat{A}_{r_\alpha}\right) = \left(\hat{A}^E, \hat{\mathbf{A}}^\mathbf{p}, \hat{A}^t, \hat{\mathbf{A}}^\mathbf{r}\right) . \tag{3.7}$$

Since we are transforming the Hamiltonian through the energy-independent substitution, then

$$\hat{A}^E \equiv 0 \ . \tag{3.8}$$

The gauge potential (3.6) leads to the substitution of the ordinary derivatives by covariant ones similarly to Eq. (2.8). As a result, by analogy with Eqs. (2.10) and (2.11), we obtain the Hamiltonian

$$\hat{\tilde{H}} = \hat{H}_d\left(\mathbf{P} - \hbar\hat{\mathbf{A}}_\mathbf{r}, \ \mathbf{R} + \hbar\hat{\mathbf{A}}_\mathbf{p}, \ t\right) - \hbar\hat{A}_t - E = \hat{\tilde{H}}_d\left(P^\alpha - \hbar\hat{A}_{r_\alpha}, \ R^\alpha + \hbar\hat{A}_{p_\alpha}\right) \ , \tag{3.9}$$

where, within the limits of Hamiltonian (3.1), the energy operator $E = i\hbar\partial/\partial t$ is responsible for the appearance of term $-\hbar\hat{A}_t$ in Eq. (3.9), when passing to the covariant derivatives. The generalized coordinates and momentums are expressed in Eq. (3.9) as

$$P^\alpha = p^\alpha + \hbar\hat{A}_{r_\alpha} \ , \quad R^\alpha = r^\alpha - \hbar\hat{A}_{p_\alpha} \ . \tag{3.10}$$

The energy value in Eq. (3.10) is shifted owing to the analogue of the scalar potential, $E \equiv P^0 = p^0 + \hbar\hat{A}^t = p^0 - \hbar\hat{A}_t$, while the time remains unchanged because of Eq. (3.8). Following this ideology, one can say that only the kinetic energy $p^0 = E_{\text{kin}}$ is a directly measured energy, while the presence of the scalar potential replaces it in the Hamiltonian with the 'generalized' (total) energy that is shifted by its potential part. Thus, the introduced 8-dimensional (in fact, 7-dimensional, Eq. (3.8)) spin gauge potential, taking into account Eq. (3.8), is in complete correspondence with the one introduced in the previous Section.

Analogously to Eq. (2.12), we determine the antisymmetric tensor of the spin gauge field:

$$\hat{F}^{\alpha\beta} = \frac{\partial\hat{A}^\beta}{\partial m_\alpha} - \frac{\partial\hat{A}^\alpha}{\partial m_\beta} - i\left[\hat{A}^\alpha, \hat{A}^\beta\right] \equiv 0 \ . \tag{3.11}$$

In the adiabatic case (see Eqs. (2.13), (2.14)) we have:

$$\hat{A}^\alpha_{(ad)} = \mathrm{dg}\,\hat{A}^\alpha = i\,\mathrm{dg}\left(\hat{U}^{-1}\frac{\partial\hat{U}}{\partial m_\alpha}\right) , \tag{3.12}$$

$$\hat{F}^{\alpha\beta}_{(ad)} = \frac{\partial\hat{A}^\beta_{(ad)}}{\partial m_\alpha} - \frac{\partial\hat{A}^\alpha_{(ad)}}{\partial m_\beta} = i\,\mathrm{dg}\left(\frac{\partial\hat{U}^{-1}}{\partial m_\alpha}\frac{\partial\hat{U}}{\partial m_\beta} - \frac{\partial\hat{U}^{-1}}{\partial m_\beta}\frac{\partial\hat{U}}{\partial m_\alpha}\right) \neq 0 \ . \tag{3.13}$$

Based on the above-said, all the energy components of $\hat{F}^{\alpha\beta}_{(ad)}$ are equal to zero:

$$\hat{F}^{\alpha E}_{(ad)} = \hat{F}^{E\alpha}_{(ad)} \equiv 0 \ . \tag{3.14}$$

Notice here that condition (2.17) of the adiabatic approximation for the particle propagating freely can be also naturally written in the 4-dimensional form:

$$\varepsilon = \frac{\hbar}{\delta p^2}\left|\frac{\partial p^\alpha}{\partial r_\beta}\right| << 1 \ , \tag{3.15}$$

where $\delta p$ stands for the distance to the nearest adjacent value of the 4-momentum with a fixed value of the 4-coordinates.

**3.3. Commutation relations and motion equations.** The adiabatic commutation relations, similar to Eqs. (2.18), are readily written in a 4-dimensional form, which at once accounts for temporal components:

$$\left[p^\alpha, p^\beta\right] = i\hbar^2\hat{F}^{(ad)}_{r_\alpha r_\beta} \ , \quad \left[r^\alpha, r^\beta\right] = i\hbar^2\hat{F}^{(ad)}_{p_\alpha p_\beta} \ , \quad \left[r^\alpha, p^\beta\right] = i\hbar g^{\alpha\beta} - i\hbar^2\hat{F}^{(ad)}_{p_\alpha r_\beta} \ , \tag{3.16}$$

where $g^{\alpha\beta} = \mathrm{diag}(-1,1,1,1)$ is the metric tensor. In line with these relations, let us write the adiabatic equations of motion in the ordinary (covariant) coordinates, as it has been done in Eqs. (2.20):

$$\dot{p}^\alpha = -\frac{i}{\hbar}\left[p^\alpha, \hat{\tilde{H}}\right] , \quad \dot{r}^\alpha = -\frac{i}{\hbar}\left[r^\alpha, \hat{\tilde{H}}\right] , \tag{3.17}$$

where the Hamiltonian (3.9) should be considered as a function of appropriate variables: $\hat{\tilde{H}} = \hat{\tilde{H}}_d(p^\alpha, r^\alpha)$. By analogy with Eqs. (2.21), in the adiabatic and semiclassical approximation these equations with commutation relations (3.16) result in the motion equation as follows:

$$\dot{p}^\alpha = -\frac{\partial \hat{\tilde{H}}_d}{\partial r_\alpha} + \hbar \hat{F}^{(ad)}_{r_\alpha m_\beta} \dot{m}_\beta \ , \quad \dot{r}^\alpha = \frac{\partial \hat{\tilde{H}}_d}{\partial p_\alpha} - \hbar \hat{F}^{(ad)}_{p_\alpha m_\beta} \dot{m}_\beta \ . \tag{3.18}$$

As we see, the commutation relations and the motion equations are expressed in the four-dimensional notation in a rather concise form.

In the case when an external electromagnetic field is present, the 4-potential of the electromagnetic field will be added in equations (3.9) and (3.10), while in the first equations of (3.16) and (3.18) the following relevant terms arise [113]:

$$\left[ p^\alpha, p^\beta \right] = i\hbar \frac{e}{c} \mathscr{F}^{\alpha\beta} + i\hbar^2 \hat{F}^{(ad)}_{r_\alpha r_\beta} \ , \tag{3.16a}$$

$$\dot{p}^\alpha = -\frac{\partial \hat{\tilde{H}}_d}{\partial r_\alpha} + \frac{e}{c} \mathscr{F}^{\alpha\beta} \dot{r}_\beta + \hbar \hat{F}^{(ad)}_{r_\alpha m_\beta} \dot{m}_\beta \ . \tag{3.18a}$$

These equations demonstrate clearly the analogy between the action of the electromagnetic field and of the adiabatic spin gauge field. The adiabatic spin gauge field $\hat{F}^{\alpha\beta}_{(ad)}$ represents actually a set of $n$ independent analogues of the electromagnetic field (each acts on its own independent spin state). Note also that the electromagnetic field acts exclusively over the coordinate $r^\alpha$-space (therefore, its components appear only in the momentum equations), whereas the spin gauge field operates on the entire phase $m^\alpha = (p^\alpha, r^\alpha)$-space.

**3.4. Berry phase.** To obtain Berry phase in terms of equations (3.9) and (3.10), let us invoke the generalized adiabatic approximation (2.17) and (3.15) for a freely propagating particle (recall that this approximation involves the semiclassics). As is known, the phase of the adiabatic solution contains the term $-i\hbar^{-1} \int E dt$, while the phase of the semiclassical solution holds the term $i\hbar^{-1} \int \mathbf{P} d\mathbf{R}$. Both of these terms can be written in a 4-dimensional form as

$$\varphi = \hbar^{-1} \int P^\alpha dR_\alpha \ . \tag{3.19}$$

Notice that it is precisely the generalized variables that enter into the phase, since the action is a function of the generalized coordinates. By expressing phase (3.19) in terms of the directly measured values from Eqs. (3.10), we obtain

$$\begin{aligned}\varphi &= \hbar^{-1} \int P^\alpha dR_\alpha = \hbar^{-1} \int \left( p^\alpha + \hbar \hat{A}_{r_\alpha} \right) d\left( r_\alpha - \hbar \hat{A}_{p^\alpha} \right) \\ &= \hbar^{-1} \int p^\alpha dr_\alpha + \int \hat{A}_{r_\alpha} dr_\alpha - p^\alpha \hat{A}_{p^\alpha} + \int \hat{A}_{p_\alpha} dp_\alpha + O(\hbar) \\ &= \hbar^{-1} \int p^\alpha dr_\alpha - \hat{A}_{p_\alpha} p_\alpha + \int \hat{A}^\alpha dm_\alpha + O(\hbar) .\end{aligned} \tag{3.20}$$

The first term here represents the dynamic phase, the second one plays no part as a consequence of its locality (for cyclic evolution it equals zero), and the third term in Eq. (3.20) represents the geometric Berry phase in the $m^\alpha$-space. In the case of the localized particle's state, where adiabaticity condition (2.16) alone takes place, the wave function phase is specified exclusively by zero (temporal) component of Eq. (3.20). In this case we arrive at Berry phase, which was discovered in the original paper [6]. Thus the semiclassical theory suggested in the paper describes completely the origination of the Berry phase. If we invoke the analogy between the adiabatic spin gauge field and the electromagnetic field, we can say that the Berry phase is an analogue of the Dirac phase [1] that reveals itself in the Aharonov-Bohm effect [2]. Various conditions for the quantization of finite states of the particle, being determined by its phase, involve the Berry phase in accordance with Eq. (3.20) (see, for example, [54,55,102–104]).

It should be noted that although we have derived the expression for Berry phase $\int \hat{A}^\alpha dm_\alpha$ in the adiabatic approximation, this formula is apparently valid in the general case as well. Indeed, in relativistic case of the Dirac equation the similar equation for Berry phase is exact [85] as is the equation for the Dirac phase. In such case the Berry phase $\int \hat{A}^\alpha dm_\alpha$ with exact potential (3.6) is bound to determine completely the entire nontrivial spin evolution of the particle, including its efficient spin transport, as a consequence of the interference effects akin to the Aharonov-Bohm effect (see Subsection 1.3).

## 4. MAXWELL EQUATIONS FOR A SPIN GAUGE FIELD

The spin gauge potential $\hat{A}^\alpha$ made its appearance after Berry's work [6], initially as the connection in an abstract parameter space. After the topological spin transport had been detected, the spin gauge field turned out to be capable of acting efficiently on a particle as a quite real force field. The next step in the description of this field is the question: *can the spin field exist in a free state?* As we have seen, the spin gauge field, in its exact form, is a purely gauge (forceless) field; its intensity $\hat{F}^{\alpha\beta}$ is identically zero (see Eqs. (1.9), (2.12), (3.11)). However, in the adiabatic approximation, this field shows itself as a force one. Hence we direct our attention just to the adiabatic case.

Let us consider the adiabatic spin gauge field $\hat{F}^{\alpha\beta}_{(ad)}$. Owing to its analogy with the electromagnetic field, the spin analogues to the Maxwell equations follow from the above-derived equations of motion (3.17), (3.18) and commutation relations (3.16) (see the Feynman derivation of the Maxwell equations in [105]). In free space they take the form:

$$\frac{\partial \hat{F}^{\alpha\beta}_{(ad)}}{\partial m^\beta} = 0 \; , \quad \frac{\partial \hat{F}^{\alpha\beta}_{(ad)}}{\partial m^\gamma} + \frac{\partial \hat{F}^{\beta\gamma}_{(ad)}}{\partial m^\alpha} + \frac{\partial \hat{F}^{\gamma\alpha}_{(ad)}}{\partial m^\beta} = 0 \; . \qquad (4.1)$$

The particular case of these equations has been considered in [106]. Equations (4.1) are invariant with respect to the gauge transformations like Eq. (2.15) (see also Subsection 1.3). As a consequence of the diagonality of $\hat{F}^{\alpha\beta}_{(ad)}$ in the spin space, they decompose into *n* independent systems with $\mathbf{U}(1)$ gauge symmetries. The Maxwell equations in free space admit the nontrivial solutions in the form of the propagating waves transporting the momentum and the energy. Hence, similar solutions in the $m^\alpha$-space can be formally constructed from equations (4.1) for the adiabatic spin gauge field $\hat{F}^{\alpha\beta}_{(ad)}$. However, it is unknown whether these solutions have a physical meaning, since the field $\hat{F}^{\alpha\beta}_{(ad)}$ has been associated for each particle only with the points of degeneracy in its own spectrum. Because, in general, the force action of the field $\hat{F}^{\alpha\beta}_{(ad)}$ describes in fact purely phase interference effects of the forceless exact field $\hat{F}^\alpha$ (see Subsections 1.3 and 1.4), it is possible that *the waves of the field $\hat{F}^{\alpha\beta}_{(ad)}$ are associated with more complex manifestations of the purely gauge filed $\hat{F}^\alpha$ in the interaction of particles.* This is a question that is yet to be explored.

As we have seen above, the electromagnetic field and the adiabatic spin gauge field act additively. Hence, in the general case, Maxwell equations (4.1) in free space should be written for a total field tensor [113]

$$\hat{G}^{\alpha\beta} = \hbar \hat{F}^{\alpha\beta}_{(ad)} + \frac{e}{c} \mathcal{F}^{\alpha\beta} \; . \qquad (4.2)$$

## 5. EXAMPLES

Below we present a number of examples (both known and new ones) that illustrate the application of the general theory constructed above. For the sake of simplicity we will consider the particles with two spin states, namely, electrons with spin 1/2 and photons with spin 1.

**5.1. Spin 1/2 in a magnetic field.** Owing to the fact that the spin is related to the electron magnetic moment, it interacts with the electromagnetic field. In this Subsection, the simplest situation, namely, the Zeeman interaction with an external magnetic field is discussed. Let us consider the spin 1/2 behavior in an external smoothly inhomogeneous, slowly nonstationary magnetic field (an analogous problem for inhomogeneous but stationary magnetic field was considered in [21–23]). The effects associated with the induced electric fields will be ignored. Then, the Hamiltonian has the form (2.3) and equals to

$$\hat{H} = \hat{I} H_0 + \hbar\chi\hat{\boldsymbol{\sigma}}\mathcal{B} , \qquad (5.1)$$

where $H_0(\mathbf{p},\mathbf{r},t)$ is the Hamiltonian in the absence of the magnetic field, $\mathcal{B} = \mathcal{B}(\mathbf{r},t)$, $\hat{\boldsymbol{\sigma}}$ stands for the electron spin operator involving the Pauli matrices, and $\chi = e/(2mc)$ ($e$ and $m$ are the electron charge and the electron mass, respectively). The unitary matrix transforming the Hamiltonian (5.1) to a diagonal form is:

$$\hat{U} = \frac{1}{\sqrt{2\mathcal{B}}} \begin{pmatrix} \sqrt{\mathcal{B}+\mathcal{B}_z} & \sqrt{\mathcal{B}-\mathcal{B}_z} \\ \dfrac{\mathcal{B}_x + i\mathcal{B}_y}{\sqrt{\mathcal{B}+\mathcal{B}_z}} & -\dfrac{\mathcal{B}_x + i\mathcal{B}_y}{\sqrt{\mathcal{B}-\mathcal{B}_z}} \end{pmatrix} , \qquad (5.2)$$

where $\mathcal{B} = |\mathcal{B}|$. With the substitution of (2.4) with (5.2), the diagonalized spin term in (5.1) takes the form

$$\hbar\hat{H}_{1d} = \hbar\chi \begin{pmatrix} \mathcal{B} & 0 \\ 0 & -\mathcal{B} \end{pmatrix} = \pm\hbar\chi\mathcal{B} . \qquad (5.3)$$

This term is responsible for Zeeman's splitting of the spin states in the electron spectrum:

$$E = E_0 \pm \hbar\chi\mathcal{B} , \qquad (5.4)$$

where $E_0$ is the energy in the absence of the magnetic field [112]. The points where $\mathcal{B} = 0$ represent the degeneracies in the electron spectrum (5.4) (see [6] and Subsection 2.5).

If the space of the magnetic field vectors $\mathcal{B}$ is treated as a three-dimensional one, where the spin term degeneracy occurs at the origin of coordinates, then the adiabatic gauge potential (2.13) induced by substitution (5.2) is equal in this space to

$$\hat{\mathbf{A}}^{(ad)}(\mathcal{B}) = \left( \frac{\mathcal{B}_y}{2\mathcal{B}(\mathcal{B}\pm\mathcal{B}_z)}, -\frac{\mathcal{B}_x}{2\mathcal{B}(\mathcal{B}\pm\mathcal{B}_z)}, 0 \right) , \qquad (5.5)$$

where different signs correspond to two spin states in (5.3), (5.4). The calculation of the field tensor (2.14) in the $\mathcal{B}$-space, which corresponds to the potential given by (5.5), yields

$$\hat{\mathbf{F}}^{(ad)}(\mathcal{B}) = \mp\frac{1}{2\mathcal{B}^3} \begin{pmatrix} 0 & \mathcal{B}_z & -\mathcal{B}_y \\ -\mathcal{B}_z & 0 & \mathcal{B}_x \\ \mathcal{B}_y & -\mathcal{B}_x & 0 \end{pmatrix} , \text{ or } \hat{F}_{ij}^{(ad)}(\mathcal{B}) = \mp\varepsilon_{ijk}\frac{\mathcal{B}_k}{2\mathcal{B}^3} . \qquad (5.6)$$

This result is consistent with the general one (2.22). We see that the force field specified is an analogue of the Dirac monopole located at the origin of coordinates in the $\mathcal{B}$-space. In this case, the efficient electron 'spin charge' equals $\pm 1/2$, i.e. the electron spin projection values.

By going from the $\mathcal{B}$-space into the phase **m**-space through the general formulas (2.23) and (2.24), we obtain the components of the adiabatic spin gauge field tensor:

$$\hat{F}_{r_i r_j}^{(ad)} = \mp\frac{\mathcal{B}}{2\mathcal{B}^3}\left( \frac{\partial\mathcal{B}}{\partial r_i} \times \frac{\partial\mathcal{B}}{\partial r_j} \right) , \quad \hat{F}_{r_i t}^{(ad)} = \mp\frac{\mathcal{B}}{2\mathcal{B}^3}\left( \frac{\partial\mathcal{B}}{\partial r_i} \times \frac{\partial\mathcal{B}}{\partial t} \right) , \quad \hat{F}_{tt}^{(ad)} = 0 . \qquad (5.7)$$

Or, in a 4-dimensional form:

$$\hat{F}^{(ad)}_{r_\alpha r_\beta} = \mp \frac{\mathscr{B}}{2\mathscr{B}^3}\left(\frac{\partial \mathscr{B}}{\partial r_\alpha} \times \frac{\partial \mathscr{B}}{\partial r_\beta}\right). \quad (5.7a)$$

In the case under consideration the $\hat{\mathbf{F}}^{(ad)}$ tensor does not possess momentum components, since the Hamiltonian spin part depends exclusively on coordinates and on time. The spatial components of the $\hat{\mathbf{F}}^{(ad)}_{\mathbf{rr}}$ tensor represent an analogue of the magnetic field tensor, whereas the spatial-temporal components of $\hat{\mathbf{F}}^{(ad)}_{\mathbf{r}t}$ present an analogue of the electric field. These values are properly involved in the motion equations (2.21) and (3.18). As a result, *the electron is moving as if it is under the action of the efficient electromagnetic field.*

$$\mathscr{F}^{\alpha\beta}_{\text{eff}} = \mathscr{F}^{\alpha\beta} + \frac{\hbar c}{e}\hat{F}^{r_\alpha r_\beta}_{(ad)} = \mathscr{F}^{\alpha\beta} \mp \frac{\hbar c}{e}\frac{\mathscr{B}}{2\mathscr{B}^3}\left(\frac{\partial \mathscr{B}}{\partial r_\alpha} \times \frac{\partial \mathscr{B}}{\partial r_\beta}\right). \quad (5.8)$$

The efficient magnetic and electric fields, in their turn, are equal, respectively, to

$$\mathscr{B}_{k\,\text{eff}} = \mathscr{B}_k \mp \frac{\hbar c}{e}\varepsilon_{ijk}\frac{\mathscr{B}}{2\mathscr{B}^3}\left(\frac{\partial \mathscr{B}}{\partial r_j} \times \frac{\partial \mathscr{B}}{\partial r_k}\right), \quad \mathscr{E}_{k\,\text{eff}} = \mp \frac{\hbar c}{e}\frac{\mathscr{B}}{2\mathscr{B}^3}\left(\frac{\partial \mathscr{B}}{\partial r_k} \times \frac{\partial \mathscr{B}}{\partial t}\right), \quad (5.9)$$

(see Remark II). Thus, the resulting adiabatic behavior of two spin states of the electron in a smoothly inhomogeneous nonstationary magnetic field can be described in terms of the charge $e$ motion in the efficient electromagnetic fields (5.8), (5.9) and the motion of the magnetic moment $\pm \hbar e / 2mc$ in the external magnetic field $\mathscr{B}$ (see (5.3)).

In the case considered by Berry [6], where the magnetic field depends on time only ($\mathscr{B} = \mathscr{B}(t)$), the force field tensor $\hat{\mathbf{F}}^{(ad)}$ on the phase space is an identical zero, and the only arising effect is the Berry phase appearance. Curious properties are revealed when one analyzes an inhomogeneous magnetic field $\mathscr{B} = \mathscr{B}(\mathbf{r})$. The point of degeneracy $\mathscr{B} = 0$ acts in the $\mathscr{B}$-space as an analogue of the magnetic monopole (see (5.6)), whereas in the real $\mathbf{r}$-space, the points of degeneracy $\mathbf{r}^*$, where $\mathscr{B}(\mathbf{r}^*) = 0$, cannot be analogous to the magnetic monopole, since the magnetic field of $\mathscr{B} \propto (\mathbf{r} - \mathbf{r}^*)$ configuration cannot exist, or in other words, by virtue of the absence of the real magnetic monopoles. This means that the degeneracy point of the Hermitian Hamiltonian (5.1) in the $\mathbf{r}$-space is not here the degeneracy point of the generic case (see [6–8]).

**5.2. Spin-orbit interaction with an electric field.** One further important case of the spin interaction with an external electromagnetic field is the spin-orbit interaction. Let us now consider it along with the Zeeman interaction. Let us place an electron in external electric and magnetic fields, which are slow functions of coordinates and of time. Then, with the spin-orbit interaction taken into account, the Hamiltonian of the electron takes the form (see (2.3)):

$$\hat{H} = \hat{\mathbf{I}}H_0 + \hbar\hat{\boldsymbol{\sigma}}\left[\chi\mathscr{B} + \rho(\mathscr{E} \times \mathbf{p})\right], \quad (5.10)$$

where the notation adopted in the previous Subsection has been retained, $\mathscr{E} = \mathscr{E}(\mathbf{r},t)$, and $\rho = e/(4m^2c^2)$.

To perform the calculations, we use the general formulas (2.22)–(2.24). For the Hamiltonian in (5.10) we have

$$\mathbf{H}_1 = \chi\mathscr{B} + \rho(\mathscr{E} \times \mathbf{p}). \quad (5.11)$$

In the $\mathbf{H}_1$-space, the adiabatic spin gauge field tensor is equal to

$$\hat{F}^{(ad)}_{ij}(\mathbf{H}_1) = \mp\varepsilon_{ijk}\frac{H_{1k}}{2H_1^3}. \quad (5.12)$$

In this case the diagonalized spin part of the Hamiltonian equals (see (5.3)):

$$\hat{H}_{1d} = \pm\hbar\left|\chi\mathscr{B} + \rho(\mathscr{E} \times \mathbf{p})\right|, \quad (5.13)$$

while the energy levels split as $E = E_0 \pm \hbar\left|\chi\mathscr{B} + \rho(\mathscr{E} \times \mathbf{p})\right|$ [112].

By using Eqs. (2.23), (2.24), the adiabatic spin gauge field tensor in the $\mathbf{m} = (\mathbf{p}, \mathbf{r}, t)$-space can be written as

$$\hat{F}^{(ad)}_{p_i p_j} = \mp \varepsilon_{klm}\varepsilon_{lpi}\varepsilon_{mrj} \frac{\rho^2(\chi\mathcal{B} + \rho(\mathcal{E}\times\mathbf{p}))_k \mathcal{E}_p \mathcal{E}_r}{2|\chi\mathcal{B} + \rho(\mathcal{E}\times\mathbf{p})|^3} = \mp\varepsilon_{klm}\varepsilon_{lpi}\varepsilon_{mrj} \frac{\chi\rho^2 \mathcal{B}_k \mathcal{E}_p \mathcal{E}_r}{2|\chi\mathcal{B} + \rho(\mathcal{E}\times\mathbf{p})|^3}, \quad (5.14a)$$

$$\hat{F}^{(ad)}_{r_i r_j} = \mp \frac{\chi\mathcal{B} + \rho(\mathcal{E}\times\mathbf{p})}{2|\chi\mathcal{B} + \rho(\mathcal{E}\times\mathbf{p})|^3}\left[\left(\chi\frac{\partial\mathcal{B}}{\partial r_i} + \rho\frac{\partial\mathcal{E}}{\partial r_i}\times\mathbf{p}\right)\times\left(\chi\frac{\partial\mathcal{B}}{\partial r_j} + \rho\frac{\partial\mathcal{E}}{\partial r_j}\times\mathbf{p}\right)\right] \quad (5.14b)$$

$$\hat{F}^{(ad)}_{p_i r_j} = \mp\varepsilon_{klm}\varepsilon_{lpi}\frac{\rho(\chi\mathcal{B} + \rho(\mathcal{E}\times\mathbf{p}))_k \mathcal{E}_p}{2|\chi\mathcal{B}+\rho(\mathcal{E}\times\mathbf{p})|^3}\left[\chi\frac{\partial\mathcal{B}}{\partial r_j} + \rho\left(\frac{\partial\mathcal{E}}{\partial r_j}\times\mathbf{p}\right)\right]_m, \quad (5.14c)$$

$$\hat{F}^{(ad)}_{p_i t} = \mp\varepsilon_{klm}\varepsilon_{lpi}\frac{\rho(\chi\mathcal{B} + \rho(\mathcal{E}\times\mathbf{p}))_k \mathcal{E}_p}{2|\chi\mathcal{B}+\rho(\mathcal{E}\times\mathbf{p})|^3}\left[\chi\frac{\partial\mathcal{B}}{\partial t} + \rho\left(\frac{\partial\mathcal{E}}{\partial t}\times\mathbf{p}\right)\right]_m, \quad (5.14d)$$

$$\hat{F}^{(ad)}_{r_i t} = \mp \frac{\chi\mathcal{B} + \rho(\mathcal{E}\times\mathbf{p})}{2|\chi\mathcal{B}+\rho(\mathcal{E}\times\mathbf{p})|^3}\left[\left(\chi\frac{\partial\mathcal{B}}{\partial r_i} + \rho\frac{\partial\mathcal{E}}{\partial r_i}\times\mathbf{p}\right)\times\left(\chi\frac{\partial\mathcal{B}}{\partial t} + \rho\frac{\partial\mathcal{E}}{\partial t}\times\mathbf{p}\right)\right] \quad (5.14e)$$

These formulas completely describe the field $\hat{\mathbf{F}}^{(ad)}$ tensor, and hence, the dynamics of a particle in this field as well. Consequently, *the adiabatic particle motion consists of the motion of the charge and the magnetic moment in the external electromagnetic field (these motions are specified by the Hamiltonian components $H_0$ and $\hbar\hat{H}_{1d}$ (5.13), respectively) along with the motion of the spin eigenstate in the $\hat{\mathbf{F}}^{(ad)}$ field (5.14)*. By analogy with (5.7a), expressions (5.14) can be represented in the 8-dimensional form in the $m^\alpha$-space.

Let us pay attention to the rather simple form of the momentum components $\hat{\mathbf{F}}^{(ad)}_{\mathbf{pp}}$ (5.14a). They determine the spin field in the momentum $\mathbf{p}$-space even with the constant fields $\mathcal{E}, \mathcal{B} = const$. As it is seen from (5.14a), $\hat{\mathbf{F}}^{(ad)}_{\mathbf{pp}}$ is nonzero only if the electric and magnetic fields exist simultaneously. To be more precise, for $\mathcal{B} = 0$, $\hat{\mathbf{F}}^{(ad)}_{\mathbf{pp}}$ has a singularity of the delta-function type at the origin of the $\mathbf{p}$-space [47,57,58]. Besides, $\hat{\mathbf{F}}^{(ad)}_{\mathbf{pp}} = 0$ with $\mathbf{p} \neq 0$, if the stationary electric and magnetic fields are mutually orthogonal.

Let us consider the case when constant electric and magnetic fields are directed at an angle to each other. Let the electric field be aligned with the $z$-axis. Then, from (5.14a) it follows that only the $\hat{F}^{(ad)}_{p_x p_y} = -\hat{F}^{(ad)}_{p_y p_x}$ components are nonzero:

$$\hat{F}^{(ad)}_{p_x p_y} = \mp\frac{\chi\rho^2\mathcal{B}_z \mathcal{E}^2}{2|\chi\mathcal{B} + \rho(\mathcal{E}\times\mathbf{p})|^3}, \quad (5.15)$$

or in the general case,

$$\hat{F}^{(ad)}_{p_i p_j} = \mp\varepsilon_{ijk}\frac{\chi\rho^2(\mathcal{B}\mathcal{E})\mathcal{E}_k}{2|\chi\mathcal{B}+\rho(\mathcal{E}\times\mathbf{p})|^3} \quad \text{or} \quad \mathbf{F}^{(ad)} = \mp\frac{\chi\rho^2(\mathcal{B}\mathcal{E})\mathcal{E}}{2|\chi\mathcal{B}+\rho(\mathcal{E}\times\mathbf{p})|^3}. \quad (5.15a)$$

Here, by analogy with (2.22), we have written the antisymmetric tensor $\hat{\mathbf{F}}^{(ad)}_{\mathbf{pp}}$ in the form of a pseudovector $\mathbf{F}^{(ad)}$. This pseudovector is an analogue of the magnetic field in the momentum $\mathbf{p}$-space and causes an additional shift, which is perpendicular both to the electric field and the external force experienced by the electron. By substituting (5.15a) into motion equations (2.21), we obtain that this shift is described by an additional term $\pm\hbar\dfrac{\chi\rho^2(\mathcal{B}\mathcal{E})}{2|\chi\mathcal{B}+\rho(\mathcal{E}\times\mathbf{p})|^3}(\dot{\mathbf{p}}\times\mathcal{E})$ in the equation for $\dot{\mathbf{r}}$. Two examples below demonstrate the influence of this shift on the particle's dynamics.

**5.3. Rashba spin-orbit interaction and the spin Hall effect.** Let us consider an electron gas in a two-dimensional solid-state lattice ($xy$-plane) with a constant magnetic field directed across this lattice (along $z$). A constant electric field is applied in the plane of the lattice (along $x$). The relativistic spin-orbit interaction discussed in the above Subsection is, as a rule, small in solids, and alternative types of the spin-orbit interactions are efficiently introduced. The simplest one is Rashba spin-orbit interaction [107], which differs from the above relativistic interaction only in the change of the fixed value $\rho$ and the replacement of the electromagnetic field by the unit vector $\mathbf{e}_z$ of the lattice symmetry (it is aligned with $z$). As a result, instead of the Hamiltonian (5.10), we have

$$\hat{H} = \hat{I} H_0 + \hbar \hat{\boldsymbol{\sigma}}[\chi \mathcal{B} + \rho(\mathbf{e}_z \times \mathbf{p})], \tag{5.16}$$

where $H_0 = \mathbf{p}^2/2m^* + e\mathcal{A}^0$ for a free electron ($m^*$ is the effective electron mass and $\mathcal{A}^0$ is the scalar potential of the electromagnetic field). Performing the substitution $\mathcal{E} \to \mathbf{e}_z$ in (5.13), (5.15) and (5.15a), and having the semiclassical motion equations (2.21), (2.21a) written, we arrive at the motion equations for a free electron:

$$\dot{\mathbf{p}} = e\mathcal{E} + \frac{e}{c}(\dot{\mathbf{r}} \times \mathcal{B}), \tag{5.17a}$$

$$\dot{\mathbf{r}} = \frac{\mathbf{p}}{m^*} \pm \hbar \frac{(\chi \mathcal{B} + \rho \mathbf{e}_z \times \mathbf{p}) \times \rho \mathbf{e}_z}{|\chi \mathcal{B} + \rho \mathbf{e}_z \times \mathbf{p}|} \pm \hbar \frac{\chi \rho^2 (\mathcal{B} \mathbf{e}_z)}{2|\chi \mathcal{B} + \rho(\mathbf{e}_z \times \mathbf{p})|^3} (\dot{\mathbf{p}} \times \mathbf{e}_z). \tag{5.17b}$$

Here the second term in the right-hand side of Eq. (5.17b) occurs as a result of the differentiation of the diagonalized spin portion of the Hamiltonian analogous to (5.13), whereas the last term in Eq. (5.17b) describes the action of the adiabatic spin gauge field $\hat{\mathbf{F}}_{\mathbf{pp}}^{(ad)}$.

It is readily seen that the second term in Eq. (5.17b) is directed along $\mathbf{p}$ and is barely noticeable against the background of the principal motion due to the semiclassics (it is proportional to $\varepsilon \sim \hbar$). Let us substitute $\dot{\mathbf{p}}$ from Eq. (5.17a) into Eq. (5.17b). The term that is proportional to $(\dot{\mathbf{r}} \times \mathcal{B}) \times \mathbf{e}_z$ is aligned with $\dot{\mathbf{r}}$, that is, almost with $\mathbf{p}$ (with an accuracy of $\hbar^2$) and also is barely noticeable against the background of the principal motion. Omitting the above-mentioned insignificant summands and taking into consideration that $\mathcal{B}\mathbf{e}_z = \mathcal{B}$, we arrive at the following equation:

$$\dot{\mathbf{r}} = \frac{\mathbf{p}}{m^*} \pm \hbar \frac{e\chi \rho^2 \mathcal{B}}{2|\chi \mathcal{B} + \rho(\mathbf{e}_z \times \mathbf{p})|^3} (\mathcal{E} \times \mathbf{e}_z). \tag{5.18}$$

Here the second term in the right-hand side is associated with the action of the $\hat{\mathbf{F}}_{\mathbf{pp}}^{(ad)}$ field and causes the additional shift of the particle across the principal current (this current is directed along the applied electric field). The particle's deviations are of opposite signs for two spin states of an electron. Thus, for an ensemble of electrons, the shift in question produces *the spin current*, which is directed *orthogonally* to the applied electric field (see [47,57–84]). In the case, when the electrons in the ensemble are polarized (one state prevails), this spin current clearly produces *the electric current* as well. Equation (5.18) describes *the anomalous and intrinsic spin Hall effects* in various magnets: the polarization of the electrons there is conditioned by the external magnetic field as well as by the sample magnetization [56–65,68].

Upon integrating equations like (5.17b) and (5.18) with respect to time, we have

$$\mathbf{r} = \mathbf{r}_0 + m^{*-1} \int \mathbf{p} \; dt - \hbar \int \hat{\mathbf{F}}_{\mathbf{pp}}^{(ad)} d\mathbf{p} = \mathbf{r}_0 + m^{*-1} \int \mathbf{p} \; dt + \hbar \int \mathbf{F}^{(ad)} \times d\mathbf{p}. \tag{5.19}$$

The longitudinal displacement caused by the term similar to (5.13) is disregarded here. As we can see, the last term in (5.19) represents a contour integral in the $\mathbf{p}$-space and has topological nature. In the example considered this term describes the topological spin splitting, transport, or pumping. If the Berry phase is given by a contour integral of the spin gauge potential $\hat{\mathbf{A}}$, then *the* *displacement during the topological adiabatic spin transport is expressed through a contour integral of the field tensor* $\hat{\mathbf{F}}^{(ad)}$.

**5.4. Topological spin transport of photons (the optical Magnus effect).** An interesting example of the spin gauge field action and the topological spin transport occurs at the propagation of light in a smoothly inhomogeneous isotropic medium in the geometrical optics approximation [101]. The geometrical optics approximation corresponds to the semiclassical and adiabatic one in quantum mechanics. Two spin states of photons correspond to the states with helicity of $\pm 1$, i.e. with right-hand and left-hand circular polarizations. The presence of the geometric Berry phase of photons reveals itself as the evolution of the polarization of light during its propagation. The evolution of this kind had been described by S. M. Rytov and V. V. Vladimirsky back in 1938 [108,109,10,101] and more recently served as the first experimental verification of the Berry phase effect (see in [9,10,13]). At a later time B.Ya. Zel'dovich et. al. discovered the displacement of light rays of distinct polarizations in the opposite directions [110], which was given the title 'the optical Magnus effect'. The subsequent phenomenological theory [111] has assigned this effect to a spin-orbit interaction. And finally, the recently developed [87–91] complete theory of geometrical optics of smoothly inhomogeneous isotropic media, which includes both the Berry phase and the optical Magnus effect, has allowed one to see the commonness of these phenomena.

Let us consider the propagation of a ray of an electromagnetic wave (photons) in a smoothly inhomogeneous medium, characterized by the refractive index $n(\mathbf{r})$, in the geometrical optics approximation. Let $\omega$ be the wave frequency, $\mathbf{k}$ be the wave vector, and $\mathbf{p} = \mathbf{k}/k_0$ be its dimensionless momentum ($k_0^{-1} = c/\omega$ is a small parameter of geometrical optics playing the role of the Planck constant in quantum mechanics). Note that the dispersion of light has the form of $\omega = \pm ck/n$. Since the frequency determines here the wave energy (dispersion of photons $E = \pm cp/n$), then at $p = k = 0$ the intersection point of terms, or the degeneracy, exists. Consequently, in the adiabatic (geometrical optics) approximation, this point generates in the $\mathbf{p}$-space the gauge potential $\hat{\mathbf{A}}_\mathbf{p}$, which corresponds to the following field tensor

$$\hat{F}^{(ad)}_{p_i p_j} = \mp \varepsilon_{ijk} \frac{p_k}{p^3} \quad \text{or} \quad \mathbf{F}^{(ad)}(\mathbf{p}) = \mp \frac{\mathbf{p}}{p^3} \ . \tag{5.20}$$

This tensor is derived from the general formulas like (2.22) for a particle with spin of 1 and with the values of helicity as the spin projection of a relativistic particle. The signs in (5.20) correspond to the helicity values of $\pm 1$.

The gauge potential $\hat{\mathbf{A}}_\mathbf{p}$ associated with the field given by (5.20) is entered into the geometrical-optics wave Hamiltonian [101] according to the general scheme (see (2.10), (2.11)):

$$H(\mathbf{p},\mathbf{R}) = \frac{1}{2}\left[p^2 - n^2\left(\mathbf{R} + k_0^{-1}\hat{\mathbf{A}}_\mathbf{p}\right)\right] = 0 \ , \tag{5.21}$$

where $\mathbf{R} = \mathbf{r} - k_0^{-1}\hat{\mathbf{A}}_\mathbf{p}$ are the generalized coordinates of the wave. The Hamiltonian ray equations for Hamiltonian (5.21) with (5.20) take the form

$$\dot{\mathbf{p}} = \frac{1}{2}\frac{\partial n^2}{\partial \mathbf{r}} \ , \quad \dot{\mathbf{r}} = \mathbf{p} + k_0^{-1}\left(\mathbf{F}^{(ad)} \times \dot{\mathbf{p}}\right) = \mathbf{p} \mp k_0^{-1}\left(\frac{\mathbf{p}}{p^3} \times \dot{\mathbf{p}}\right) \ . \tag{5.22}$$

(In these equations the dot signifies the differentiation with respect to the parameter associated with the ray length [87–90,101].) The last term in the second equation (5.22) is responsible for the displacement of a ray across its direction depending on the sign of the helicity. The identical summand has been introduced by Zel'dovich and Lieberman in [111] to describe the optical Magnus effect. The rigorous detailed derivation for these equations in the framework of the theory of Berry phase and spin gauge fields was performed in [86–91].

Thus we can say that *during the propagation of photons in an inhomogeneous medium, the spin current is generated, which is perpendicular to the propagation direction and is similar in* *nature to the intrinsic spin Hall effect*. This results in the topological spin (polarization) *splitting* of the rays of different helicities, which is observed experimentally. The description of this effect along with examples and illustrations can be found in [88,90,91].

## 6. SUMMARY

In this paper we have considered the evolution of particle with a spin. This particle is described by $n$-dimensional matrix Hamiltonian. In the general case this Hamiltonian depends on coordinates, momentum and time. In the first Section the basic ideas, leading to the introduction of the spin gauge field along with the Berry phase, have been detailed. The spin gauge potential is a pure gauge one and results from the Hamiltonian diagonalizing transformation from the group $\mathbf{U}(n)$. This transformation allows one to bring the Hamiltonian, in the general case, into the diagonalized form and *to treat the evolution of various spin states of a particle as the evolution of the spin multiplet of scalar particles in a spin gauge field*. Thus, it is precisely this non-Abelian field that describes all transitions between various states. Since each normalized Hamiltonian eigenvector is determined with an accuracy of an arbitrary transformation $\mathbf{U}(1)$, the entire gauge field is gauge-invariant with respect to the group $\mathbf{U}^n(1) \subset \mathbf{U}(n)$. That is, the spin gauge field, *being a pure gauge one with respect to the $\mathbf{U}(n)$ group, is gauge-invariant with respect to the $\mathbf{U}^n(1)$ group*, which allows developing the gauge-invariant theory relating to this group.

The spin gauge field intensity in the exact (nonadiabatic) case is identical zero (a forceless field). However, because the spin gauge potential is non-Abelian, nontrivial phase interference effects may occur during the particle evolution, which are associated with the spin gauge potential. First of all, the particle's phase contains a summand akin to the Dirac phase in the electromagnetic field that is well-studied and is referred to as the Berry phase. The non-Abelian phase factor corresponding to the Berry phase contains the whole nontrivial 'topological' evolution of spin states of a particle. In addition to the Aharonov-Bohm type interference effect, the phase effects resulting in the displacement of particle beams like the Shelankov effect [3] may also occur. These are just the manifestations of the topological spin transport of particles, which is now of particular interest.

In our paper no consideration has been given to the mentioned general interference effects associated with the exact non-Abelian Berry phase factor. Basically, the adiabatic evolution of a particle has been discussed. In this case, the spin gauge potential can be approximately replaced with the diagonal (Abelian) one, which gives rise to non-zero field tensor. Thus the adiabatic spin gauge field becomes a force field, and its action on a particle can be described in the ordinary way through the motion equations. It should be stressed that the transition to the adiabatic case does no more than replaces a purely phase action of the exact forceless field with its approximate efficient force analogue. In view of the diagonality of the adiabatic equations and fields, they actually decompose into $n$ independent components, each describing the evolution of an independent spin state. At the same time, each component of the adiabatic spin gauge field represents an analogue of the electromagnetic field on the particle's phase space. This analogy, together with the motion equations that involve the forces of the adiabatic spin gauge field action, comprises the main portion of the present investigation. Different spin states of a particle represent effectively different charges in an adiabatic spin gauge field. This causes the difference in the forces acting on different spin states, and, as a consequence, the topological spin transport of particles (or topological spin pumping and splitting). In addition to the general theory, we have presented in this paper a number of specific examples, both known and new ones.

The main results that are obtained in the present paper for the first time are as follows. 1) We have derived in the Hamiltonian formalism general motion equations of a particle with a spin that take into account the adiabatic spin gauge field on the generalized phase space. 2) It is pointed out that in exact (i.e. without adiabatic reduction) case the evolution of a particle can be presented as the evolution in a non-Abelian pure gauge field with non-Abelian analogue of the Aharonov-Bohm effect, which results in topological spin transport of the particle. 3) It is shown that the adiabatic spin gauge field represents a complete analogue of the electromagnetic field (previously the existence of analogues of magnetic field only were reported (see Remark II). 4) Owing to the complete analogy with electromagnetic field the theory is reduced to a 4-dimensional form. 5) The hypothesis of the existence of the field equation for the adiabatic spin gauge field is presented (see also [106]). 6) Finally, an important example of adiabatic and semiclassical motion of spin 1/2 in external inhomogeneous and non-stationary electric and magnetic fields is considered in details with description of the spin gauge field action on the generalized phase space.

In conclusion, let us list the principal questions that, in our opinion, should first of all be considered in order to develop the complete adequate theory of the spin gauge fields and the evolution of a particle with spin.

1. In a sense the spin can be assigned to manifestations of relativistic phenomena. Hence the question arises as to the natural introduction of a spin gauge field in the framework of relativistic equations. In particular, in the Dirac equation the Berry phase as well as the Dirac phase has exact and not the adiabatic character [85]. With the exact expression for the Berry phase factor, we can describe exactly all effects associated with the spin gauge field as well.

2. One principal current problem is to describe the topological spin transport effects in the context of the exact (forceless) non-Abelian spin gauge field. As already noted, in our opinion, these effects should be described through the exact Berry phase factor as the non-Abelian analogues of the Aharonov-Bohm effect and Shelankov's beam deflections [3]. In terms of non-Abelian gauge fields, the case in point is the description of the Wilson loop associated with the spin gauge field.

3. An interesting question is the possibility of the existence and wave propagation of spin gauge fields (possibly, in the adiabatic approximation) in the free state. This problem is posed in Section 4 of the present paper.

The next step will hopefully be a theory involving the outstanding questions listed above.

## ACKNOWLEDGMENTS

The authors are sincerely grateful to S.A. Gredeskul, K.V. Il'enko, K.A. Kikoin and Yu.P. Stepanovskiy for fruitful discussions and their attention to the work. The work was supported in part by INTAS (Grant No. 03-55-1921), Ukrainian President's Grant for Young Scientists GP/F8/51, and the Center for Absorption in Science of the Ministry of Immigrant Absorption of Israel.